\documentclass[11pt]{article}

\PassOptionsToPackage{authoryear}{natbib}

\usepackage[authoryear]{natbib}
\usepackage[margin=1.0in]{geometry}

\usepackage{xcolor}
\usepackage[utf8]{inputenc}
\usepackage[T1]{fontenc}    
\usepackage[hidelinks]{hyperref}
\usepackage{url}            
\usepackage{booktabs}       
\usepackage{amsfonts}      
\usepackage{nicefrac}       
\usepackage{microtype}     
\usepackage{enumitem}
\usepackage{amsmath}
\usepackage{bbm}
\usepackage{graphicx}
\usepackage{subcaption}
\usepackage{datetime2}
\usepackage{setspace}
\usepackage{rotating}
\usepackage{comment}
\usepackage{listings}
\usepackage{multirow}
\usepackage{algorithm}
\usepackage{algorithmic}

\usepackage{newtxtext}
\usepackage{newtxmath}
\DeclareMathSizes{5}{5}{3.5}{2.5}

\makeatletter
\let\old@makefntext\@makefntext
\renewcommand{\@makefntext}[1]{\setstretch{1}\old@makefntext{#1}}
\makeatother

\title{\textbf{Evaluating A/B Testing Methodologies via \\ Sample Splitting: Theory and Practice}}
\date{}

\author{%
\begin{tabular}{ccc}
   Ryan Kessler\thanks{Email: \texttt{rykessle@amazon.com}, \texttt{jmcq@amazon.com}, \texttt{miikka@amazon.com}. We thank Dmitry Taubinsky for inspiring discussions and Thomas Richardson and James Robins for helpful comments and suggestions.} & James McQueen & Miikka Rokkanen \\
   \emph{Amazon} & \emph{Amazon} & \emph{Amazon} \\
\end{tabular}
}

\begin{document}

\maketitle

\begin{abstract}
\begin{spacing}{0.95}
We develop a theoretical framework for sample splitting in A/B testing environments, where data for each test are partitioned into two splits to measure methodological performance when the true impacts of tests are unobserved. We show that sample-split estimators are generally biased for full-sample performance but consistently estimate sample-split analogues of it. We derive their asymptotic distributions, construct valid confidence intervals, and characterize the bias-variance trade-offs underlying sample-split design choices. We validate our theoretical results through simulations and provide implementation guidance for A/B testing products seeking to evaluate new estimators and decision rules. 
\end{spacing}
\end{abstract}

\clearpage

\section{Introduction}

Measuring the impact of proposed improvements to the way in which A/B tests are analyzed and used to inform decision making is critical for large-scale A/B testing products. Over the last few decades, there have been several proposed improvements to A/B testing methodologies, including new estimators of impacts \citep[e.g.,][]{deng_etal_2013, deng_etal_2023, deng_etal_2024} and new rules for deciding when to launch \citep[e.g.,][]{goldberg_johndrow_2017, azevedo_etal_2019, azevedo_etal_2020}. The scope of proposed improvements continues to expand, with more recent work exploring AI-driven approaches \citep[e.g.,][]{delriochanona_etal_2023, delriochanona_etal_2025}. A/B testing products must determine which of these methodologies deliver meaningful value relative to their implementation costs.

The key challenge is that the true impacts of A/B tests are unobservable.  Each test yields only a noisy estimate of the true impact, making it difficult to tell whether a proposed methodology actually leads to an overall improvement in estimates or decision making across tests. While simulation methods can be used to evaluate performance, they require strong assumptions about the true underlying data generating process. 

Sample splitting offers a potential solution to this problem \citep{tripuraneni_etal_2023, parikh_etal_2026}, but its theoretical foundation remains underdeveloped. Under sample splitting, each A/B test is divided into two non-overlapping splits: one to implement the proposed methodology, the other to construct an independent, unbiased proxy for the true impact. Comparing these two quantities across A/B tests allows us to measure the average performance of the proposed methodology without making strong distributional assumptions \citep{tripuraneni_etal_2023, parikh_etal_2026}. However, the statistical properties of these sample-split estimators have yet to be characterized formally: their bias for full-sample performance, variance, and asymptotic distributions remain unknown, and there are no general methods for conducting statistical inference. Without such foundations, products lack guidance on exactly how to use sample splitting, limiting its adoption in practice.

In this paper, we introduce a theoretical framework for sample splitting in A/B testing environments, characterizing both what these methods can measure and how to conduct valid inference. We show that while sample splitting enables performance evaluation when true impacts are unobserved, it does so at a cost. Because methodologies are implemented on only a split of the data, sample-split estimators are generally biased for what we ultimately care about --- how the methodologies perform on the full sample. We show that sample-split estimators consistently estimate sample-split analogues of full-sample performance, derive their asymptotic distributions, and show how to construct valid asymptotic confidence intervals. We demonstrate that while a careful choice of how the sample is split can reduce bias for full-sample performance, this improvement generally comes at the cost of increased variance. We validate our main theoretical results through simulations. 

Our theoretical framework has important practical implications for A/B testing products. First, it shows that, in some cases, products can dramatically reduce the computational costs of sample splitting by simulating sample-split estimates rather than repeatedly partitioning unit-level data. Second, it provides a principled approach to statistical inference, enabling products to conduct hypothesis testing (e.g., to determine whether one methodology significantly outperforms another) and statistical power calculations (e.g., to determine how many A/B tests are needed to reliably detect a given performance difference between methodologies). Third, it demonstrates that common sample splitting practices  --- such as splitting data equally or using hundreds of random partitions --- are often suboptimal. These insights, together with the step-by-step algorithms we provide, enable products to implement principled sample splitting efficiently at scale, making it a practical and powerful tool for evaluating proposed methodological improvements.

More generally, our work contributes to the literature on cross-validation and resampling methods \citep{stone_1971}. Like cross-validation, sample splitting separates the application of a methodology from its evaluation by partitioning data. The key difference is that in cross-validation the prediction target is observed directly, whereas in our setting the prediction target is itself an estimate subject to sampling variation. To our knowledge, \citet{tripuraneni_etal_2023} were the first to apply the idea of sample splitting in the A/B testing context, building upon earlier work in the estimation of heterogeneous treatment effects by \citet{athey_imbens_2016} and \citet{chernozhukov_2025}. \citet{tripuraneni_etal_2023} demonstrate the potential of sample splitting in evaluating the performance of A/B testing methodologies but leave its statistical properties largely unformalized.\footnote{The authors acknowledge that their work is guided primarily by intuition and heuristics rather than principled theory. For example, they note the use ``$t$-test heuristics'' (page 7) and suggest that the confidence intervals they report ``should be interpreted cautiously'' (page 11).} After the first version of this paper was posted, \citet{parikh_etal_2026} developed formal results for cross-fitted comparisons of estimators under mean squared error and regret and proposed permutation-based inference. Because the estimators in their cross-fitting procedure are applied to training folds, their analysis leaves unaddressed the gap between split-sample and full-sample performance. We formalize this distinction, analyze its bias-variance implications, and develop asymptotic inference for split-sample performance.

\section{Motivating Theory}
\label{sec:theory}

\subsection{Context}
\label{sec:context}

We consider a set of $i \in \{1,...,I\}$ feature changes, each evaluated through an A/B test where the change is exposed to a random subset of units. Each feature change has a true impact $\Delta_i$, which is a random variable with realized values $\delta_i$. Each test is further characterized by latent variables $Z_i$, with realized values $z_i$, capturing features of the testing environment such as sample size and outcome variability. The pairs $(\Delta_i, Z_i)$ are independently and identically distributed according to an unknown distribution $G$. 

Conditional on the true impact $\Delta_i = \delta_i$ and test environment $Z_i = z_i$, the A/B test generates random data $\hat{X}_i$ according to a sampling distribution $F_i$, with realized values $\hat{x}_i$. To simplify notation, we suppress the dependence on $i$ and write $F$ in place of $F_i$. This sampling distribution reflects the sampling of units and random assignment of treatment in each test and may vary across tests through the true impact and test environment $(\Delta_i, Z_i)$.  

We define an \emph{A/B testing methodology} $\kappa(\hat{x}_i)$ to be a mapping of the data $\hat{x}_i$ to a real-valued output. The methodology could be an estimator of the true impact $\delta_i$ or a decision rule that determines whether to launch the given treatment ($\kappa(\hat{x}_i) = 1$) or not ($\kappa(\hat{x}_i) = 0$). 

Given the randomization of the feature change and the central limit theorem, we assume that each test yields an unbiased estimator $\hat{\Delta}_i$ of the impact $\delta_i$ that is normally distributed, with realized values $\hat{\delta}_i$ and sampling variance $\tau_i^2 = \tau_i^2(\delta_i, z_i)$: 

\begin{equation}
   \label{eq:likelihood}
   \hat{\Delta}_i |\Delta_i = \delta_i, Z_i = z_i \overset{\text{ind}}{\sim}  N(\delta_i, \tau_i^2)
\end{equation}

Following standard practice in the A/B testing literature \citep[see, e.g.,][]{azevedo_etal_2020}, we treat $\tau_i^2$ as known for each test, while allowing it to vary across tests.

We measure the performance of the methodology $\kappa(\hat{x}_i)$ given a true impact of $\Delta_i = \delta_i$ via a real-valued \emph{performance function} $p(\kappa(\hat{x}_i), \delta_i)$. As illustrated in the examples below, the specification of the performance function depends on the type of methodology being evaluated. 

\vspace{0.5pc}

\noindent \textbf{Example 1 (Measuring performance of an estimator):} We may want to measure the performance of a shrinkage estimator $\kappa(\hat{x}_i) = w_i \cdot \hat{\delta}_i$ for $w_i \in (0, 1)$ under squared error in which case we could specify the performance function $p(\kappa(\hat{x}_i), \delta_i) = (\kappa(\hat{x}_i) - \delta_i)^2$.

\vspace{0.5pc}

\noindent \textbf{Example 2 (Measuring performance of a decision rule):} We may want to measure the performance of the decision rule $\kappa(\hat{x}_i) = 1(\hat{\delta}_i > \tau_i \cdot \Phi^{-1}(0.975))$ to launch treatments when the estimated impact is positive and statistically significant at the 5 percent level. In this case, we could specify the performance function $p(\kappa(\hat{x}_i), \delta_i) = \kappa(\hat{x}_i) \cdot \delta_i - (1 - \kappa(\hat{x}_i)) \cdot \delta_i$ to capture the idea that we get $\delta_i$ when we launch and $-\delta_i$ when we do not. We hereafter refer to this measure of performance and its counterpart $p(\kappa(\hat{x}_i), \delta_i) = \kappa(\hat{x}_i) \cdot \delta_i$ as \emph{decision value} and \emph{launch-only decision value}, respectively.

\vspace{0.5pc}

In what follows, for any random variable $K_i$, we write $F_K$ to denote its distribution induced by the sampling distribution $F$ and $G_K$ to denote its distribution induced by the distribution $G$. We write $E_F(\cdot)$ and $\text{Var}_F(\cdot)$ for the expectation and variance with respect to randomness induced by $F$, and $E_G(\cdot)$ and $\text{Var}_G(\cdot)$ for the expectation and variance with respect to randomness induced by $G$. Finally, we write $E(\cdot)$ and $\text{Var}(\cdot)$ for the expectation and variance with respect to both sources of randomness, integrating over both $F$ and $G$. We denote by $\phi(\cdot)$ and $\Phi(\cdot)$ the probability density function and cumulative distribution function of the standard normal distribution, respectively. 

\subsection{Ideal Estimands}
\label{sec:estimands}

Our goal is to quantify how well different methodologies perform across A/B tests. To this end, we begin by defining a set of \emph{ideal estimands} that capture how we would measure performance if the true impacts $\delta_i$ were observable. These estimands serve as our conceptual targets and help clarify the sense in which our estimators (introduced in section \ref{sec:estimators_inference}) may be biased for what we ultimately want to measure. 

For each A/B test $i$, the expected performance of a methodology $\kappa$ under performance measure $p$ given a true impact of $\Delta_i = \delta_i$ and test environment $Z_i = z_i$ is given by: 

\begin{equation}
    \label{eq:ideal_estimand_eq1}
    Y_i(\kappa) = E_F\left(p\left(\kappa\left(\hat{X}_i\right), \delta_i \right)\right) 
\end{equation}

Here $\kappa(\hat{X}_i)$ is the methodology's output (estimate or launch decision) given data $\hat{X}_i$. The function $p(\kappa(\hat{X}_i), \delta_i)$ measures how well this output performs given a true impact $\delta_i$. Taking the expectation over the sampling variability induced by $F$ averages out the randomness in the data $\hat{X}_i$, yielding the expected performance $Y_i(\kappa)$ for a given true impact and test environment. 

While $Y_i(\kappa)$ captures the expected performance of $\kappa$ for a given test, it varies across tests due to variation in $(\Delta_i, Z_i)$:

\begin{equation}
    \label{eq:ideal_estimand_eq2}
    Y_i(\kappa) \overset{\text{iid}}{\sim} G_Y, \: \text{with} \: E_{G}\left(Y_i(\kappa)\right) = \theta(\kappa), \text{Var}_{G}\left(Y_i(\kappa)\right) =\lambda^2(\kappa)
\end{equation}

\vspace{0.5pc}

One ideal estimand is therefore the average performance of the methodology across tests, $\theta(\kappa)$. The quantity $\theta(\kappa)$ accounts for both the sampling variation in each A/B test and the variation in true impacts across tests, making it a natural target when evaluating methodological performance at scale. 

While average performance can be helpful in benchmarking a given methodology, we are often more interested in differences in average performance across methodologies. Another ideal estimand is therefore:

\begin{equation}
    \label{eq:ideal_estimand_eq3}
    \theta(\kappa^1, \kappa^2) = \frac{\theta(\kappa^2) -  \theta(\kappa^1)}{\theta(\kappa^1)}
\end{equation}

\noindent where $\kappa^1$ and $\kappa^2$ are different methodologies, and we normalize by $\theta(\kappa^1)$ so that $\theta(\kappa^1, \kappa^2)$ can be interpreted as the percent change in performance of methodology $\kappa^2$ relative to $\kappa^1$.

\vspace{0.5pc}

\noindent \textbf{Example 3 (Illustrating performance estimands for estimators):} Assume that $G$ is such that $\tau_i = \tau$ for all $i=1,..., I$ and $\Delta_i \sim N(0, \sigma^2)$. Consider evaluating the relative performance of two estimators under squared error $p(\kappa(\hat{x}_i), \delta_i) = (\kappa(\hat{x}_i) - \delta_i)^2$: the unbiased estimator $\kappa^1(\hat{x}_i)=\hat{\delta}_i$ and the Bayes estimator $\kappa^2(\hat{x}_i) = \sigma^2 (\sigma^2 + \tau^2)^{-1} \hat{\delta}_i$. In this case, we have that:

\begin{equation}
    \theta(\kappa^1, \kappa^2) = \frac{\frac{\sigma^2 \tau^2}{\sigma^2 + \tau^2} - \tau^2}{\tau^2} = \frac{\sigma^2}{\sigma^2 + \tau^2} - 1 < 0
\end{equation}

\noindent consistent with the fact that the Bayes estimator minimizes mean squared error \citep{casella_berger_2002}.

\vspace{0.5pc}

\noindent \textbf{Example 4 (Illustrating performance estimands for decision rules):} Assume that $G$ is such that $\tau_i = \tau$ for all $i=1,..., I$ and $\Delta_i \sim N(0, \sigma^2)$. Consider evaluating the relative performance of two decision rules under launch-only decision value $p(\kappa(\hat{x}_i), \delta_i) = \kappa(\hat{x}_i) \cdot \delta_i$: a frequentist decision rule $\kappa^1(\hat{x}_i)= 1(\hat{\delta}_i > \tau \cdot \Phi^{-1}(0.975))$ which launches treatments when the estimated impact is positive and statistically significant at the 5 percent level and a Bayes decision rule $\kappa^2(\hat{x}_i) = 1(\sigma^2 (\sigma^2 + \tau^2)^{-1} \hat{\delta}_i > 0)$ which launches treatments when the posterior mean is positive. In this case, we have that:

\begin{equation}
    \theta(\kappa^1, \kappa^2) = \frac{\frac{\sigma^2}{\sqrt{\sigma^2+\tau^2}} \left(\phi(0)  - \phi\left(\frac{\tau \cdot \Phi^{-1}(0.975)}{\sqrt{\sigma^2 + \tau^2}}\right) \right)}{\frac{\sigma^2}{\sqrt{\sigma^2+\tau^2}} \phi\left(\frac{\tau \cdot \Phi^{-1}(0.975)}{\sqrt{\sigma^2 + \tau^2}}\right)}  =  \frac{\phi(0)}{\phi\left(\frac{\tau \cdot \Phi^{-1}(0.975)}{\sqrt{\sigma^2 + \tau^2}}\right)} - 1 > 0
\end{equation}

\noindent consistent with the fact that the Bayes decision rule maximizes mean launch-only decision value \citep{azevedo_etal_2020}.

\vspace{0.5pc}

Note that these examples make restrictive assumptions about the distribution $G$ only to help make the estimands concrete. Our sample splitting framework, introduced below, is designed to evaluate performance of methodologies without requiring any assumptions about $G$. 

\subsection{Sample Splitting}

We evaluate the performance of methodologies via sample splitting, where we partition each A/B test into two splits: one to implement the proposed methodology, the other to construct an independent, unbiased proxy for the true impact. This approach enables performance measurement when true impacts are unobserved \citep{tripuraneni_etal_2023}. However, because the proposed methodology is implemented on only a split of the sample, we cannot directly estimate the ideal estimands in section \ref{sec:estimands}. In what follows, we outline a general approach to generating sample-split data, which applies to any methodology. We then show that this approach can be simplified for ``plug-in methodologies'' that depend on the data only through a set of sufficient statistics. With this data foundation, we then define sample-split analogues of the ideal estimands, introduce corresponding unbiased estimators, and outline an approach to statistical inference.

\subsubsection{Data Generation: General Approach}
\label{sec:data_general}

We implement sample splitting as follows. For each A/B test, we randomly partition the units assigned to each experience (control and treatment) into two splits, labeled $a$ and $b$, with a fraction $\alpha \in (0, 1)$ allocated to split $a$ and the remaining $(1-\alpha)$ to split $b$. For a given random partition $s$, let $\hat{X}_{ias}$ and $\hat{X}_{ibs}$ be the data associated with units in splits $a$ and $b$, with $(\hat{x}_{ias} \cup \hat{x}_{ibs}) = \hat{x}_{i}$ and $(\hat{x}_{ias} \cap \hat{x}_{ibs}) = \emptyset$. We use split $a$ for \emph{training}, applying both the methodology of interest and the unbiased estimator $\hat{\Delta}_i$ of $\delta_i$. This yields an estimate or launch decision $\kappa(\hat{x}_{ias})$ and an estimate $\hat{\delta}_{ias}$. We use split $b$ for \emph{evaluation}, applying the unbiased estimator $\hat{\Delta}_i$ of $\delta_i$ to obtain a proxy for the true impact $\hat{\delta}_{ibs}$. Because we treat $\tau_i^2$ as known, we do not estimate sampling variances of $\hat{\Delta}_{ias}$ and $\hat{\Delta}_{ibs}$ based on data but instead compute them deterministically based on $\tau_i^2$ and the split fraction $\alpha$. Specifically, $\tau_{ias}^2 = \tau_i^2/\alpha$ and $\tau_{ibs}^2 = \tau_i^2 / (1 - \alpha)$, reflecting the standard inverse relationship between sampling variance and sample size. 

Because the methodology $\kappa(\hat{X}_{ias})$ and the proxy for the true impact $\hat{\Delta}_{ibs}$ are based on disjoint samples of units, they are statistically independent conditional on the true impact $\Delta_i = \delta_i$ and testing environment $Z_i = z_i$: 

\begin{equation}
    \label{eq:sample_split_independent}
    \kappa\left(\hat{X}_{ias}\right) \perp \hat{\Delta}_{ibs} \: \big| \:\Delta_i = \delta_i, Z_i = z_i 
\end{equation}

Additionally, because the partitions are determined at random, the proxy for the true impact $\hat{\Delta}_{ibs}$ is (under equation (\ref{eq:likelihood})) unbiased and normally distributed conditional on the true impact $\Delta_i = \delta_i$: 

\begin{equation}
    \label{eq:sample_split_unbiased}
    \hat{\Delta}_{ibs}  | \Delta_i = \delta_i, Z_i = z_i \overset{\text{ind}}{\sim} N\left(\delta_i,  \tau_{ibs}^2 \right)
\end{equation}

Together, equations (\ref{eq:sample_split_independent}) and (\ref{eq:sample_split_unbiased}) provide the statistical foundation for our sample splitting methodology. The conditional independence property in equation (\ref{eq:sample_split_independent}) ensures that noise affecting the methodology's output does not also influence its validation, guarding against spurious estimates of performance. The normality and unbiasedness properties in equation (\ref{eq:sample_split_unbiased}) ensure that split $b$ provides a valid proxy for the true impact $\delta_i$.

A single partition suffices to estimate performance, but results can vary substantially across different partitions. To reduce this variability, we repeat our sample splitting procedure $S$ times, generating new random partitions that each divide units into splits $a$ and $b$. This yields estimates $\{\kappa(\hat{x}_{ias}), \hat{\delta}_{ias}, \tau_{ias}^2, \hat{\delta}_{ibs}, \tau_{ibs}^2\}_{s=1}^S$ which we then average for each test in hopes of obtaining more stable results. Importantly, while splits $a$ and $b$ remain independent within each partition $s$ (maintaining the property in equation (\ref{eq:sample_split_independent})), they are not independent across partitions, as they are based on overlapping data. Our approach to estimation and inference, outlined in section \ref{sec:estimators_inference}, takes this dependence into account. 

\subsubsection{Data Generation: Plug-in Methodologies}
\label{sec:data_plug_in}

While the previous section illustrated how sample-split data can be generated by repeatedly partitioning the unit-level data of a test, many methodologies depend on the unit-level data only through the sufficient statistics $(\hat{\delta}_{ias}, \tau_{ias}^2)$. We refer to these as ``plug-in methodologies'', as they simply plug the estimates  $(\hat{\delta}_{ias}, \tau_{ias}^2)$ into deterministic functions, without changing the way in which the estimates themselves are constructed. Plug-in methodologies include shrinkage estimators $\kappa(\hat{x}_{ias}) = w_i \cdot \hat{\delta}_{ias}$ for $w_i \in (0, 1)$ and threshold decision rules $\kappa(\hat{x}_{ias}) = 1(\hat{\delta}_{ias} > \tau_{ias} \cdot c)$ for launch threshold $c \geq 0$. 

For these methodologies, we can simplify sample-split data generation. Instead of repeatedly partitioning unit-level data and re-computing estimates, we can directly sample $(\hat{\Delta}_{ias}, \hat{\Delta}_{ibs})$ from their joint distribution given the full-sample estimates $(\hat{\delta}_i, \tau_i^2)$ and true impact $\delta_i$. This shortcut, formalized in the proposition below, eliminates the need for unit-level data in partitioning while preserving the statistical guarantees that sample splitting provides in performance evaluation. 

\vspace{0.5pc}

\noindent \textbf{Proposition 1 (Evaluating plug-in methodologies without unit-level partitioning)}: Let splits $a$ and $b$ be formed by randomly partitioning units in an A/B test with proportions $\alpha$ and $(1-\alpha)$, respectively. Under equation (\ref{eq:likelihood}), the joint distribution of $(\hat{\Delta}_{ias}, \hat{\Delta}_{ibs})$ conditional on the full-sample estimates $(\hat{\delta}_i, \tau_i^2)$ and true impact $\delta_i$ is given by:

\begin{equation}
\label{eq:prop1}
\begin{pmatrix}
\hat{\Delta}_{ias} \\
\hat{\Delta}_{ibs}
\end{pmatrix}
\Big |
\hat{\Delta}_i = \hat{\delta}_i, \Delta_i = \delta_i, \tau_i^2
\sim N
\begin{pmatrix}
\begin{pmatrix}
\hat{\delta}_i \\
\hat{\delta}_i
\end{pmatrix}
,
\begin{pmatrix}
\frac{\tau_i^2}{\alpha} - \tau_i^2 & -\tau_i^2 \\
-\tau_i^2 & \frac{\tau_i^2}{1-\alpha} - \tau_i^2
\end{pmatrix}
\end{pmatrix}
\end{equation}

\noindent For a given test, the process of repeatedly partitioning units and re-estimating $(\hat{\delta}_{ias}, \hat{\delta}_{ibs})$ can therefore be approximated by resampling from the conditional distribution in equation (\ref{eq:prop1}) given the full-sample estimates $(\hat{\delta}_i, \tau_i^2)$. Given draws from this distribution, plug-in methodologies (being deterministic functions of $(\hat{\delta}_{ias}, \tau_{ias}^2)$) can then be applied, yielding estimates $\{\kappa(\hat{x}_{ias}), \hat{\delta}_{ias}, \tau_{ias}^2, \hat{\delta}_{ibs}, \tau_{ibs}^2\}_{s=1}^S$.\footnote{Equation (\ref{eq:prop1}) implies that, conditional on full-sample estimates $(\hat{\delta}_i, \tau_i^2)$, $\hat{\Delta}_{ias}$ and  $\hat{\Delta}_{ibs}$ are perfectly negatively correlated across partitions. This follows from the fact that $\hat{\Delta}_i = \alpha \hat{\Delta}_{ias} + (1-\alpha) \hat{\Delta}_{ibs}$, so increases in one require decreases in the other.}

\noindent \emph{Proof:} See appendix \ref{sec:proofs} $\hfill \square$

\vspace{0.5pc}

Proposition 1 is similar in spirit to \citet{lewis_2025}, who generate ``doppelganger'' observations without access to unit-level data by deconvolving the overall estimates for a test.

By eliminating the need for unit-level data in partitioning, proposition 1 lowers the computational cost of evaluating plug-in methodologies. Without the shortcut, evaluating plug-in methodologies requires processing $N$ unit observations for each of the $I$ tests $S$ times. With it, we simply draw from the conditional distribution $S$ times for each of the $I$ tests. In contexts with millions of units and thousands of tests, the savings can be substantial, enabling us to increase the scale of our analyses by orders of magnitude and obtain significantly more precise estimates of performance. 

However, not all methodologies can be evaluated via this shortcut. The shortcut relies on the methodology treating the estimates $(\hat{\delta}_{ias}, \tau_{ias}^2)$ as sufficient statistics for the underlying data $\hat{x}_{ias}$. Some methodologies --- such as covariate adjustment, stratification, and outlier handling ---  affect how these estimates are constructed and therefore still require repeated unit-level partitioning.  

\subsubsection{Estimands} 
\label{sec:sample_split_estimands}

To evaluate performance of methodologies under sample splitting, we introduce generalized, sample-splitting analogues of the ideal estimands introduced in section \ref{sec:estimands}. These estimands reflect that performance depends not only on the methodology $\kappa$ and performance function $p$, but also on the share of the sample $\alpha$ used in implementing the methodology. Given $\alpha$, the expected performance of a methodology $\kappa$ under performance measure $p$ given a true impact of $\Delta_i = \delta_i$ and testing environment $Z_i = z_i$ is given by: 

\begin{equation}
    \label{eq:split_estimand_eq1}
    Y_i(\kappa, \alpha) = E_F\left(p\left(\kappa\left(\hat{X}_{ias}\right), \delta_i \right)\right) 
\end{equation}

This generalization nests the ideal case: as $\alpha \rightarrow 1$ the methodology is implemented on the full sample and we recover the ideal performance measure in equation (\ref{eq:ideal_estimand_eq1}). 

Just like its ideal counterpart $Y_i(\kappa)$, $Y_i(\kappa, \alpha)$ varies across tests due to variation in $(\Delta_i, Z_i)$: 

\begin{equation}
    \label{eq:split_estimand_eq2}
    Y_i\left(\kappa, \alpha \right) \overset{\text{iid}}{\sim} G_Y, \: \text{with} \: E_{G}\left(Y_i\left(\kappa, \alpha\right)\right) = \theta\left(\kappa, \alpha\right), \text{Var}_{G}\left(Y_i\left(\kappa, \alpha\right)\right) = \lambda^2\left(\kappa, \alpha\right)
\end{equation}

\noindent We denote realizations of $Y_i(\kappa, \alpha)$ by $y_i(\kappa, \alpha)$. 

As in section \ref{sec:estimands}, our sample-splitting estimands are then the average performance of the methodology across tests, $ \theta\left(\kappa, \alpha\right)$, and percent differences in average performance across methodologies: 

\begin{equation}
    \label{eq:split_estimand_eq3}
    \theta(\kappa^1, \kappa^2, \alpha) = \frac{\theta(\kappa^2, \alpha) -  \theta(\kappa^1, \alpha)}{\theta(\kappa^1, \alpha) }
\end{equation}

\vspace{0.5pc}

\noindent \textbf{Example 5 (Illustrating performance estimands for estimators, continued):} Returning to example 3, we now consider the sample-split analogue of the difference in mean squared error across the estimators. For any $\alpha < 1$, sample splitting overestimates the performance benefits of the Bayes estimator. Replacing $\tau^2$ with $\tau^2 / \alpha$, we have that:

\begin{equation}
    \theta(\kappa^1, \kappa^2, \alpha) = \frac{\sigma^2}{\sigma^2 + \frac{\tau^2}{\alpha}} - 1 <  \frac{\sigma^2}{\sigma^2 + \tau^2} - 1=  \theta(\kappa^1, \kappa^2) 
\end{equation}

\noindent For example, when $\sigma^2 = \tau^2 = 1$, the Bayes estimator reduces mean squared error by 50 percent. Sample splitting with $\alpha = 0.50$ would, however, suggest a reduction of 66 percent. 

\vspace{0.5pc}

\noindent \textbf{Example 6 (Illustrating performance estimands for decision rules, continued):}  Returning to example 4, we now consider the sample-split analogue of the difference in launch-only decision value across the decision rules. For any $\alpha < 1$, sample splitting overestimates the performance benefits of the Bayes decision rule. Replacing $\tau^2$ with $\tau^2 / \alpha$, we have that:

\begin{equation}
    \theta(\kappa^1, \kappa^2, \alpha) =  \frac{\phi(0)}{\phi\left(\frac{\sqrt{\frac{\tau^2}{\alpha}} \cdot \Phi^{-1}(0.975)}{\sqrt{\sigma^2 + \frac{\tau^2}{\alpha}}}\right)}  - 1>  \frac{\phi(0)}{\phi\left(\frac{\tau \cdot \Phi^{-1}(0.975)}{\sqrt{\sigma^2 + \tau^2}}\right)}  - 1= \theta(\kappa^1, \kappa^2)
\end{equation}

\noindent For example, when $\sigma^2 = \tau^2 = 1$, the Bayes decision rule increases the launch-only decision value by 160 percent. Sample splitting with $\alpha = 0.50$ would, however, suggest a benefit of 260 percent.

\vspace{0.5pc}

As these examples illustrate, the sample-split estimands can differ substantially from their ideal counterparts, with the magnitude of this difference governed by the split fraction $\alpha$. In subsequent sections, we discuss how to estimate these sample-split estimands and the tradeoffs underlying different choices of $\alpha$. 

\subsubsection{Estimators and Inference}
\label{sec:estimators_inference}

Sample splitting enables us to estimate the performance of methodologies by replacing the unobserved true impacts $\delta_i$ with the independent, unbiased proxies $\hat{\Delta}_{ibs}$ from split $b$. This leads to natural estimators of the sample-split estimands $Y_i(\kappa, \alpha)$ introduced in the previous section.  

For each A/B test $i$ and partition $s$, we consider an estimator $\hat{Y}_{is}(\kappa, \alpha)$ whose average across partitions $\hat{Y}_{i}(\kappa, \alpha, S)=\frac{1}{S}\sum_{s=1}^S \hat{Y}_{is}(\kappa, \alpha)$ satisfies:

\vspace{-1pc}
\begin{align}
    \label{eq:sample_split_estimator_eq1}
    \hat{Y}_{i}(\kappa, \alpha, S) | Y_i(\kappa, \alpha) = y_i(\kappa, \alpha) \overset{\text{ind}}{\sim} F_{\hat{Y}}, \: \text{with} \: E_F\left(\hat{Y}_{i}(\kappa, \alpha, S)\right) &= y_i(\kappa, \alpha), \\
     \text{Var}_F\left(\hat{Y}_{i}(\kappa, \alpha, S)\right) &= \Omega_i^2\left(\kappa, \alpha, S\right) \nonumber
\end{align}

The variance $\Omega_i^2\left(\kappa, \alpha, S\right)$ reflects both sampling variability in the full-sample data $\hat{X}_i$ given the true impact $\Delta_i = \delta_i$ and testing environment $Z_i = z_i$ and variability in the sample-split data $(\hat{X}_{ias}, \hat{X}_{ibs})$ given the full-sample data. Because the sampling distribution $F_i$ varies across tests only through $(\Delta_i, Z_i)$, $\Omega_i^2\left(\kappa, \alpha, S\right)$ is likewise determined by $(\Delta_i, Z_i)$. Since $(\Delta_i, Z_i)$ are independently and identically distributed across tests, $\hat{Y}_i(\kappa, \alpha, S)$ are therefore independently and identically distributed across tests under the superpopulation model.

Given $\hat{Y}_{i}(\kappa, \alpha, S)$ satisfying equation (\ref{eq:sample_split_estimator_eq1}), we estimate the average performance of methodologies across tests and differences in average performance across methodologies. We estimate the average performance across tests $\theta(\kappa, \alpha)$ via:

\begin{equation}
     \label{eq:sample_split_estimator_eq2}
    \hat{\theta}(\kappa, \alpha, S) = \frac{1}{I} \sum_{i=1}^I \hat{Y}_i(\kappa, \alpha, S) 
\end{equation}

We estimate the difference in average performance across methodologies $\theta(\kappa^1, \kappa^2, \alpha)$ via: 

\begin{equation}
     \label{eq:sample_split_estimator_eq3}
     \hat{\theta}(\kappa^1, \kappa^2, \alpha, S) = \frac{\hat{\theta}(\kappa^2, \alpha, S) - \hat{\theta}(\kappa^1, \alpha, S)}{\hat{\theta}(\kappa^1, \alpha, S)}
\end{equation}

Because each $\hat{Y}_{i}(\kappa, \alpha, S)$ is unbiased for $y_i(\kappa, \alpha)$, the estimator $\hat{\theta}(\kappa, \alpha, S)$ is unbiased for $\theta(\kappa, \alpha)$. We formalize this via the following proposition. 

\vspace{0.5pc}

\noindent \textbf{Proposition 2 (Unbiasedness)}: Under equations (\ref{eq:split_estimand_eq2}) and (\ref{eq:sample_split_estimator_eq1}), the estimator of average performance $\hat{\theta}(\kappa, \alpha, S)$ is unbiased for $\theta(\kappa, \alpha)$.

\noindent \emph{Proof:} See appendix \ref{sec:proofs} $\hfill \square$

\vspace{0.5pc}

The challenge is therefore to construct estimators $\hat{Y}_{is}(\kappa, \alpha)$ that satisfy equation (\ref{eq:sample_split_estimator_eq1}) for different performance functions $p$. Table \ref{tab:estimands_estimators} presents such estimators for common choices of $p$, including bias, mean squared error, and decision value. We provide proofs of the unbiasedness condition in (\ref{eq:sample_split_estimator_eq1}) for each estimator in appendix \ref{sec:proofs}, relying on the sample splitting conditions in equations (\ref{eq:sample_split_independent})  and (\ref{eq:sample_split_unbiased}). 

\begin{table}[h!]
\centering 
\caption{Sample-split estimands and unbiased estimators by performance measure}
\label{tab:estimands_estimators}
\vspace{0.5pc}
\resizebox{1.0\columnwidth}{!}{%
\begin{tabular}{llllc}
    \toprule
    Methodology & Measure of performance $p$  & Sample-split estimand $Y_i(\kappa, \alpha)$ & Unbiased estimator $\hat{Y}_{is}(\kappa, \alpha)$  \\ \midrule
    Estimator & Bias & $E_F(\kappa(\hat{X}_{ias}) - \delta_i)$  & $\kappa(\hat{x}_{ias}) - \hat{\delta}_{ibs}$\\ 
    Estimator & Squared error & $E_F((\kappa(\hat{X}_{ias})- \delta_i)^2)$ &  $(\kappa(\hat{x}_{ias}) - \hat{\delta}_{ibs})^2 - \tau_{ibs}^2$ \\ 
    Decision rule & Decision value & $E_F(\kappa(\hat{X}_{ias}) \cdot \delta_i - (1 - \kappa(\hat{X}_{ias})) \cdot \delta_i )$ &  $\kappa(\hat{x}_{ias}) \cdot \hat{\delta}_{ibs} - (1 - \kappa(\hat{x}_{ias})) \cdot \hat{\delta}_{ibs}$ \\ 
     Decision rule & Launch-only decision value & $E_F(\kappa(\hat{X}_{ias}) \cdot \delta_i)$ &  $\kappa(\hat{x}_{ias}) \cdot \hat{\delta}_{ibs}$ \\ \bottomrule
\end{tabular}%
}
\vspace{-0.1pc}
{\flushleft \tiny
\begin{spacing}{0.4}
\emph{Notes}: This table presents sample-split estimands $Y_i(\kappa, \alpha)$ and corresponding unbiased estimators $\hat{Y}_{is}(\kappa, \alpha)$ for different measures of performance. 
\end{spacing}}
\end{table}

Although $\hat{\theta}(\kappa, \alpha, S)$ is unbiased for $\theta(\kappa, \alpha)$, the estimator of the difference in average performance across methodologies $\hat{\theta}(\kappa^1, \kappa^2, \alpha, S)$ is generally biased for $\theta(\kappa^1, \kappa^2, \alpha)$. The bias arises from the nonlinearity of the percent change: even when $\hat{\theta}(\kappa^1, \alpha, S)$ and $\hat{\theta}(\kappa^2, \alpha, S)$ are unbiased for $\theta(\kappa^1, \alpha)$ and $\theta(\kappa^2, \alpha)$, respectively, their ratio $\hat{\theta}(\kappa^1, \kappa^2, \alpha, S)$ need not be unbiased for $\theta(\kappa^1, \kappa^2, \alpha)$. 

Having established the bias properties of the estimators $\hat{\theta}(\kappa, \alpha, S)$ and $\hat{\theta}(\kappa^1, \kappa^2, \alpha, S)$, we now turn to their large-sample behavior. In practice, we care not only about whether the estimators are close to their true values on average, but also about their sampling variability and the rate at which they converge to their true values as the number of tests $I$ grows large. 

We summarize our results via the two following propositions. For brevity, we focus the propositions on $\hat{\theta}(\kappa, \alpha, S)$. Provided $\theta(\kappa^1, \alpha) \neq 0$, analogous results for $\hat{\theta}(\kappa^1, \kappa^2, \alpha, S)$ follow from the multivariate central limit theorem and the delta method applied to the paired test-level performance estimates.

\vspace{0.5pc}

\noindent \textbf{Proposition 3 (Asymptotic normality)}: Under equations (\ref{eq:split_estimand_eq2}) and (\ref{eq:sample_split_estimator_eq1}) and the regularity condition that $E(\hat{Y}_i(\kappa, \alpha, S)^2) < \infty$, the estimator $\hat{\theta}(\kappa, \alpha, S)$ is asymptotically normally distributed:

\begin{equation}
    \sqrt{I}\left(\hat{\theta}(\kappa, \alpha, S) - \theta(\kappa, \alpha)\right) \overset{d}{\longrightarrow} N\left(0, \zeta^2(\kappa, \alpha, S) \right)
\end{equation}

\noindent where $\zeta^2(\kappa, \alpha, S) =  \lambda^2(\kappa, \alpha) +  E_G\left(\Omega_i^2\left(\kappa, \alpha, S\right)\right)$.

\noindent \emph{Proof:} See appendix \ref{sec:proofs} $\hfill \square$

\vspace{0.5pc}

Proposition 3 shows that the estimator $\hat{\theta}(\kappa, \alpha, S)$ is asymptotically normally distributed. Its asymptotic variance $\zeta^2(\kappa, \alpha, S)$ captures heterogeneity in true performance across tests via $\lambda^2(\kappa, \alpha)$ and the average sampling variance of estimated performance across tests via $E_G\left(\Omega_i^2\left(\kappa, \alpha, S\right)\right)$. 

Proposition 4 shows that $\zeta^2(\kappa, \alpha, S)$ can be consistently estimated, enabling valid asymptotic inference.  

\vspace{0.5pc}

\noindent \textbf{Proposition 4 (Valid asymptotic inference):} Under equations (\ref{eq:split_estimand_eq2}) and (\ref{eq:sample_split_estimator_eq1}) and the regularity condition that $E(\hat{Y}_i(\kappa, \alpha, S)^2) < \infty$, the estimator:

\begin{equation}
 \hat{\zeta}^2(\kappa, \alpha, S) = \frac{1}{I} \sum_{i=1}^I (\hat{Y}_i(\kappa, \alpha, S) - \hat{\theta}(\kappa, \alpha, S))^2
\end{equation}
 
\noindent is consistent for the asymptotic variance $\zeta^2(\kappa, \alpha, S) =  \lambda^2(\kappa, \alpha) + E_G\left(\Omega_i^2\left(\kappa, \alpha, S\right)\right)$. Therefore, the $(1-\upsilon)$ confidence intervals:

\begin{equation}
    \hat{\theta}(\kappa, \alpha, S) \pm \frac{\hat{\zeta}(\kappa, \alpha, S)}{\sqrt{I}} \cdot \Phi^{-1}\left(1-\frac{\upsilon}{2}\right) \\
\end{equation}

\noindent will cover $\theta(\kappa, \alpha)$ with probability $(1-\upsilon)$ as $I \longrightarrow \infty$.

\noindent \emph{Proof:} See appendix \ref{sec:proofs} $\hfill \square$

\vspace{0.5pc}

Together, these results enable principled estimation of the performance of methodologies via sample splitting. The unbiasedness of the estimators allows us to reliably estimate average performance under sample splitting, while the asymptotic normality enables us to quantify uncertainty in our estimates. The asymptotic confidence intervals can be used both for hypothesis testing (e.g., to determine whether one methodology significantly outperforms another) and statistical power calculations (e.g., to determine how many A/B tests are needed to reliably detect a given performance difference between methodologies). Algorithm 1 provides a step-by-step guide towards implementation.

\vspace{0.5pc}

\begin{algorithm}
\caption{Estimating average performance of a methodology $\kappa$}
{\footnotesize
\begin{enumerate}
    \item Identify $I$ historical A/B tests and choose split fraction $\alpha$ and number of partitions $S$
    \item Generate sample-split data:
    \begin{enumerate}
        \item If evaluating a plug-in methodology, simulate data, as described in section \ref{sec:data_plug_in}
        \item Otherwise, generate data via unit-level partitioning, as described in section \ref{sec:data_general}
    \end{enumerate}
    \item Choose measure of performance $\hat{Y}_{is}(\kappa, \alpha)$ from table \ref{tab:estimands_estimators} and compute average across partitions $\hat{Y}_{i}(\kappa, \alpha, S) = \frac{1}{S} \sum_{s=1}^S \hat{Y}_{is}(\kappa, \alpha)$ for each test
    \item Construct an estimate of average performance $\hat{\theta}(\kappa, \alpha, S) =\frac{1}{I} \sum_{i=1}^I \hat{Y}_i(\kappa, \alpha, S)$ and estimate its asymptotic variance via $\hat{\zeta}^2(\kappa, \alpha, S) = \frac{1}{I} \sum_{i=1}^I \left(\hat{Y}_i(\kappa, \alpha, S) - \hat{\theta}(\kappa, \alpha, S)\right)^2$
    \item Construct $(1-\upsilon)$ confidence intervals for average performance:
    \begin{equation*}
    \hat{\theta}(\kappa, \alpha, S) \pm \frac{\hat{\zeta}(\kappa, \alpha, S)}{\sqrt{I}} \cdot \Phi^{-1}\left(1-\frac{\upsilon}{2}\right)
    \end{equation*}
\end{enumerate}}
\end{algorithm}

Note, however, that care must be taken in applying these asymptotic results in smaller samples of A/B tests. The normal approximation in proposition 3 requires a sufficient number of A/B tests to be reliable. Similarly, the variance estimator in proposition 4 may be unstable in small samples, potentially leading to poor coverage of the resulting confidence intervals. Our simulation results in the next section suggest that these approximations work well with thousands of tests, but practitioners should exercise caution when working with significantly smaller samples. 

\section{Simulations}

In this section, we simulate a stylized A/B testing environment in which the difference in average performance between two methodologies is known. These simulations serve three purposes. First, they validate theoretical predictions outlined in previous sections. Second, they assess the adequacy of the asymptotic approximations underlying our approach to estimation and inference in finite samples. Third, they allow us to assess how performance of our estimators varies as a function of key sample-split design parameters. 

\subsection{Data Generating Process} 

We simulate from the following data generating process (DGP):

\begin{align}
    \Delta_i & \overset{\text{iid}}{\sim} N(0, \sigma^2) \\
    \epsilon_i &\overset{\text{iid}}{\sim} \chi_1^2 \\
    \tau_i^2 &= \tau^2 + \epsilon_i \\
    \hat{\Delta}_i | \Delta_i = \delta_i, \tau_i^2 &\overset{\text{ind}}{\sim} N(\delta_i, \tau_i^2) \\
    \begin{pmatrix}
        \hat{\Delta}_{ias} \\
        \hat{\Delta}_{ibs}
    \end{pmatrix}
    \Big | \hat{\Delta}_i = \hat{\delta}_i, \Delta_i = \delta_i, \tau_i^2 & \sim N
    \begin{pmatrix}
\begin{pmatrix}
\hat{\delta}_i \\
\hat{\delta}_i
\end{pmatrix}
,
\begin{pmatrix}
\frac{\tau_i^2}{\alpha} - \tau_i^2 & -\tau_i^2 \\
-\tau_i^2 & \frac{\tau_i^2}{1-\alpha} - \tau_i^2
\end{pmatrix}
\end{pmatrix}
\end{align}

We vary five key parameters: the variance of true impacts $\sigma^2$, baseline sampling variance $\tau^2$, split fraction $\alpha$, number of A/B tests $I$, and number of random partitions $S$. 

This DGP captures three key features of standard A/B testing environments. First, we capture heteroskedasticity across tests by modeling the sampling variance of the estimated impacts via $\tau_i^2 = \tau^2 + \epsilon_i$, where $\tau$ is a constant and $\epsilon_i$ is a chi-squared random variable with 1 degree of freedom. Second, we capture different signal-to-noise ratios by considering different pairwise combinations of the variance of true impacts $\sigma^2$ and the baseline sampling variance $\tau^2$. Third, we capture the real-world correlations across partitions $s$ for a given test by drawing $(\hat{\Delta}_{ias}, \hat{\Delta}_{ibs})$ from their joint distribution given the full-sample estimates $(\hat{\delta}_i, \tau_i^2)$.

As a benchmark specification, we consider $I = 5K$ A/B tests, $S=30$ partitions, a split fraction $\alpha = 0.5$, a variance of true impacts $\sigma^2 = 1$, and a baseline sampling variance $\tau^2 = 2$. 

\subsection{Evaluation}

We use the simulations to evaluate the relative performance of different point estimators and decision rules. For point estimators, we compare the performance of the unbiased estimator $\kappa^1(\hat{x}_i)=\hat{\delta}_i$ to that of the Bayes estimator $\kappa^2(\hat{x}_i) = \sigma^2 (\sigma^2 + \tau_i^2)^{-1} \hat{\delta}_i$ under squared error, building upon example 3. For decision rules, we compare the performance of the frequentist decision rule $\kappa^1(\hat{x}_i)= 1(\hat{\delta}_i > \tau_i \cdot \Phi^{-1}(0.975))$ to that of a Bayes decision rule $\kappa^2(\hat{x}_i) = 1(\sigma^2 (\sigma^2 + \tau_i^2)^{-1} \hat{\delta}_i > 0)$ under launch-only decision value, building upon example 4. 

For each methodology type, we measure the bias of the estimator $\hat{\theta}(\kappa^1, \kappa^2, \alpha, S)$, both relative to the ideal estimand $\theta(\kappa^1, \kappa^2)$ and the sample-split estimand $\theta(\kappa^1, \kappa^2, \alpha)$. We also measure the variance of $\hat{\theta}(\kappa^1, \kappa^2, \alpha, S)$ and the coverage of its 95 percent confidence interval. Finally, we assess how the bias and variance vary as a function of the number of tests $I$, number of partitions $S$, and split fraction $\alpha$. 

\subsection{Results}

\begin{figure}[h!]
\centering 
\caption{Bias and coverage of 95 percent confidence intervals at different $(\tau^2, \sigma^2)$}
\label{fig:simulations_main}
\begin{tabular}{ccc}
    \multicolumn{3}{c}{\scriptsize (a) Bayes estimator relative to unbiased estimator under squared error} \\ [0.25pc] 
    {\scriptsize(i) Bias for ideal estimand} & {\scriptsize (ii) Bias for sample-split estimand} & {\scriptsize (iii) Coverage of sample-split estimand}  \\ 
    \includegraphics[clip, trim=1cm 0cm 0cm 0.8cm, scale = 0.35]{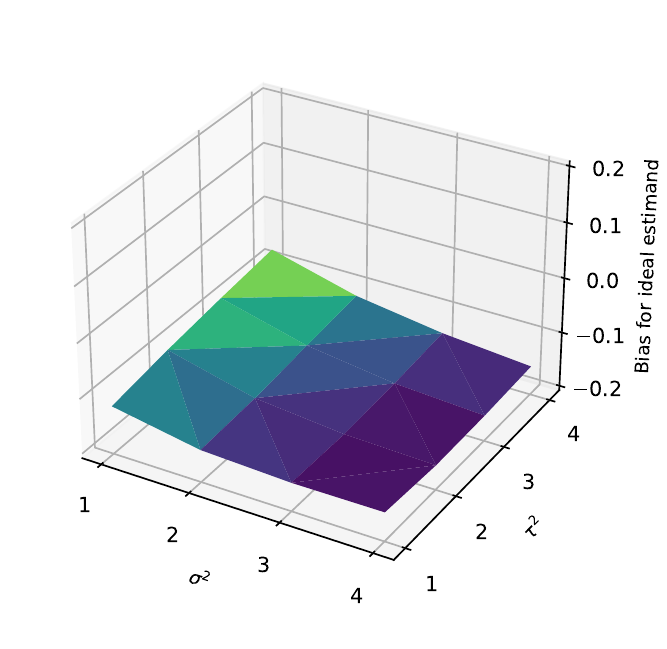} &  \includegraphics[clip, trim=1cm 0cm 0cm 0.8cm, scale = 0.35]{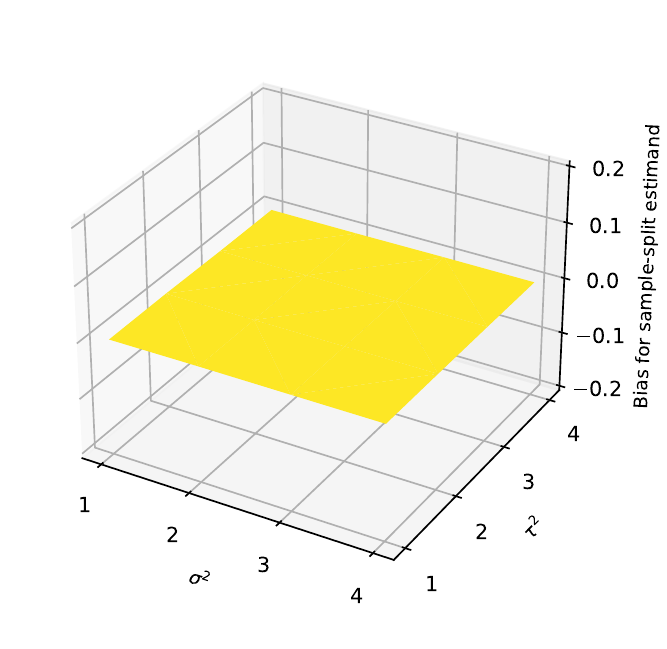} & \includegraphics[clip, trim=1cm 0cm 0cm 0.8cm, scale = 0.35]{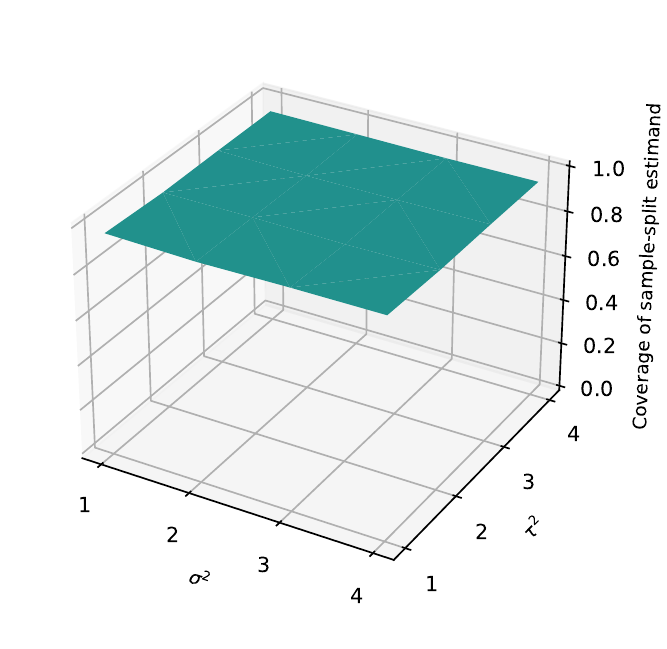} \\
    \multicolumn{3}{c}{\scriptsize (b) Bayes decision rule relative to frequentist decision rule under launch-only decision value} \\ [0.25pc] 
    {\scriptsize (i) Bias for ideal estimand} & {\scriptsize (ii) Bias for sample-split estimand} & {\scriptsize (iii) Coverage of sample-split estimand}  \\ 
    \includegraphics[clip, trim=1cm 0cm 0cm 0.8cm, scale = 0.35]{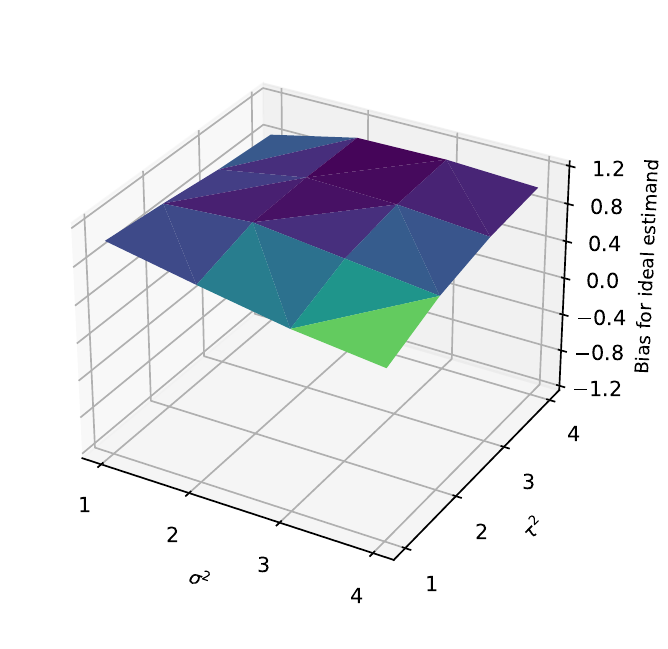} &  \includegraphics[clip, trim=1cm 0cm 0cm 0.8cm, scale = 0.35]{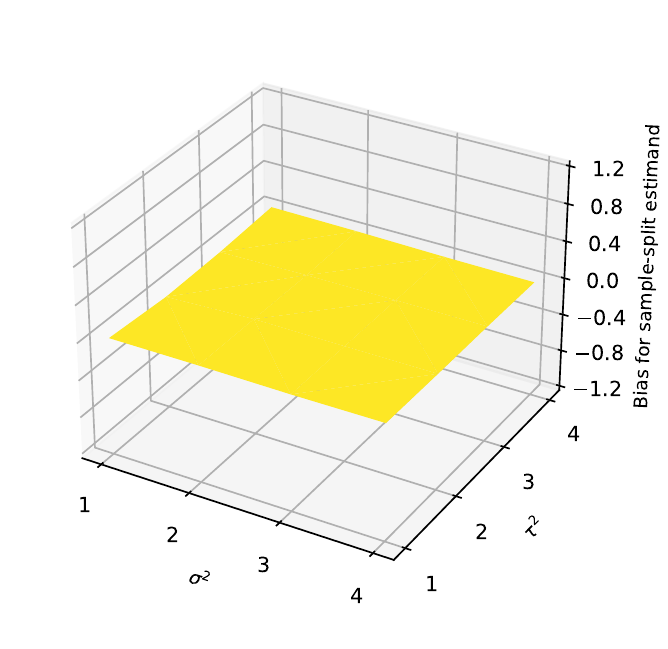} & \includegraphics[clip, trim=1cm 0cm 0cm 0.8cm, scale = 0.35]{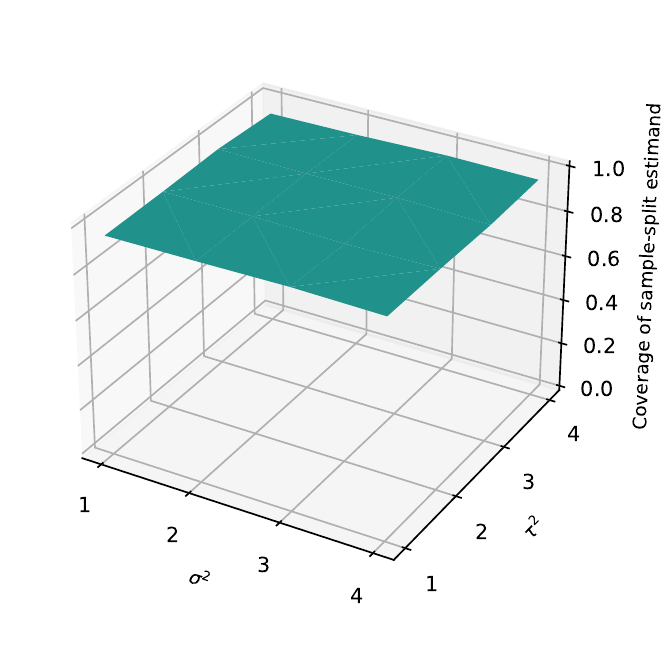} 
\end{tabular}
\end{figure}

Figure \ref{fig:simulations_main} confirms the theoretical predictions outlined in previous sections. Sample splitting estimators are biased for the ideal estimands but are consistent for the sample-split estimands, resulting in negligible bias in the sample of $I = 5K$ A/B tests. The confidence intervals achieve nominal coverage, covering the sample-split estimand approximately 95 percent of the time. 

\begin{figure}[h!]
\centering 
\caption{Bias-variance tradeoff underlying choice of split fraction $\alpha$}
\label{fig:bias_variance_alpha}
\begin{tabular}{cc}
    {\scriptsize (a)  Bayes estimator relative to unbiased } & {\scriptsize (b) Bayes decision rule relative to frequentist}  \\ [-0.25pc] 
    {\scriptsize estimator under squared error} & {\scriptsize decision rule under launch-only decision value} \\ 
    \includegraphics[scale = 0.40]{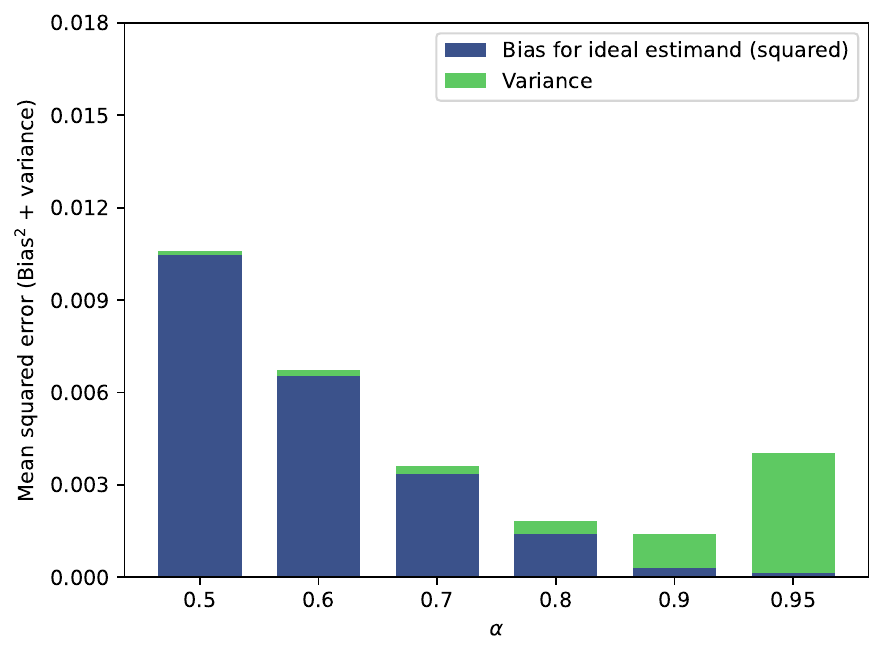} & \includegraphics[scale = 0.40]{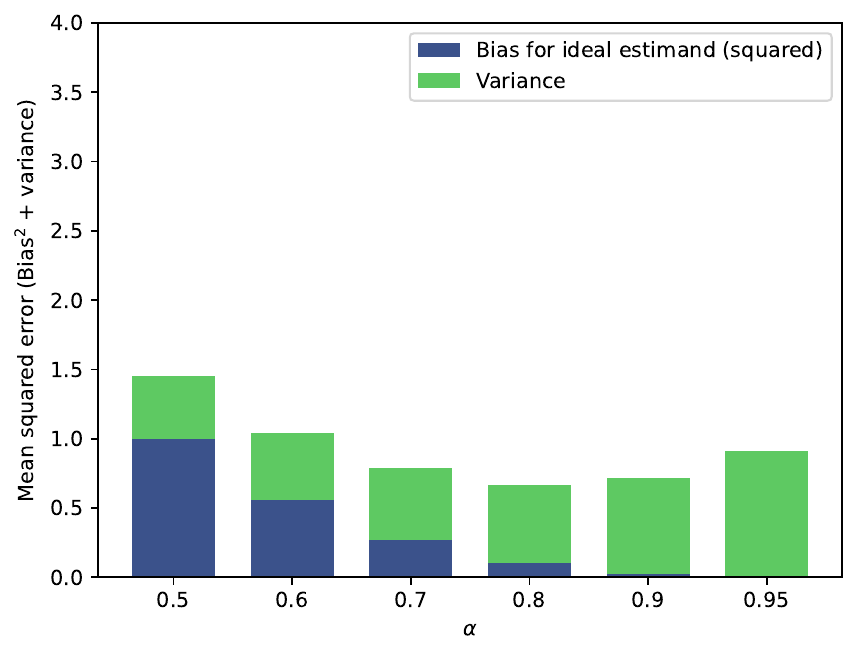}
\end{tabular}
\end{figure}

Figure \ref{fig:bias_variance_alpha} highlights a bias-variance tradeoff underlying the choice of the split fraction $\alpha$. As more data is allocated to the training split (i.e., as $\alpha \rightarrow 1$), bias for the ideal estimand falls to zero but the variance of the estimated difference in performance rises. Under the baseline DGP, this produces a $U$-shaped curve in terms of mean-squared error, with the minimizing $\alpha$ greater than 0.5. We present results for alternative specifications of the DGP in appendix \ref{sec:supplemental_results}.

Together, the results in figures \ref{fig:simulations_main} and \ref{fig:bias_variance_alpha} underscore both the value and limitations of sample splitting. Sample splitting makes estimation possible by using a split of the data to construct an independent, unbiased proxy for the true impact, thereby overcoming the fundamental challenge that true impacts are unobserved. However, because we implement the methodology on only a split of the data, we generally cannot construct unbiased estimators of the quantity we ultimately care about: the performance of the methodology if it had been implemented on the full sample. We can reduce bias for this ideal estimand by increasing $\alpha$ but this comes at the cost of increasing the variance of our estimates of performance. 

\begin{figure}[h!]
\centering 
\caption{Variance as a function of the number of tests $I$ and number of partitions $S$}
\label{fig:variance_i_s}
\begin{tabular}{cc}
    {\scriptsize (a)  Bayes estimator relative to unbiased } & {\scriptsize (b) Bayes decision rule relative to frequentist}  \\ [-0.25pc] 
    {\scriptsize estimator under squared error} & {\scriptsize decision rule under launch-only decision value} \\ 
     \includegraphics[scale = 0.40]{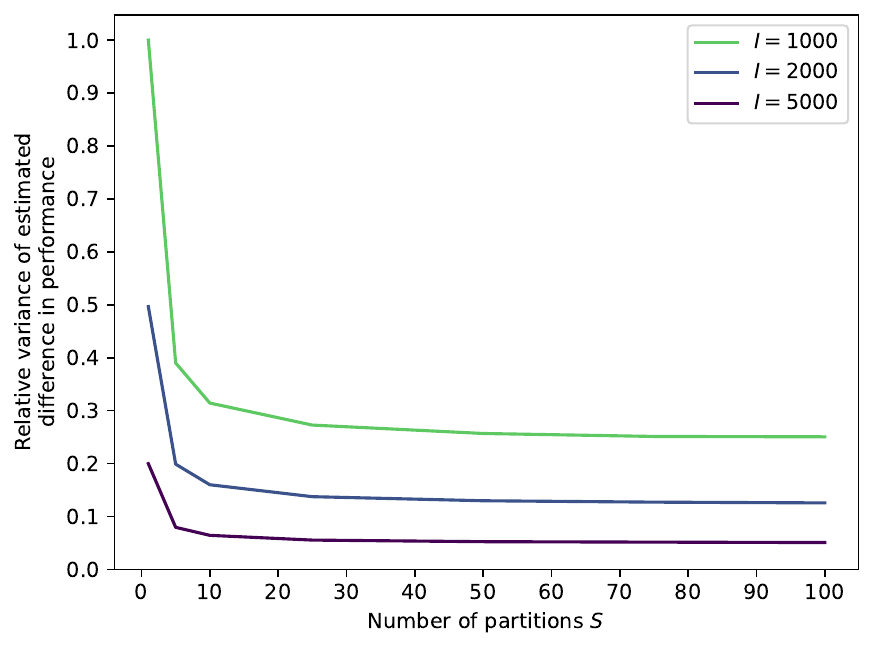} & \includegraphics[scale = 0.40]{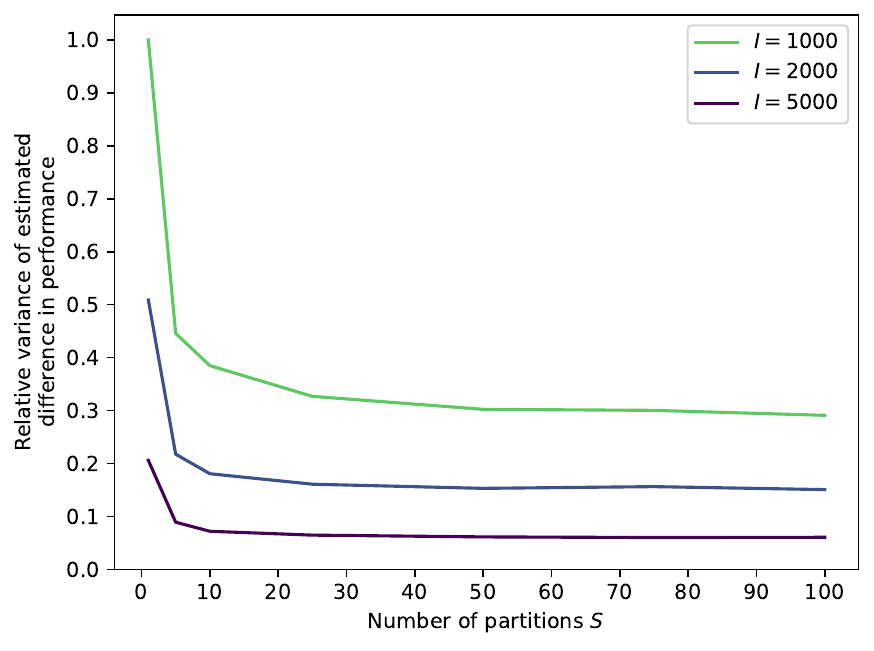}
\end{tabular}
\end{figure}

Figure \ref{fig:variance_i_s} illustrates how the variance of sample-split estimators depends on both the number of A/B tests $I$ and partitions $S$. While increasing either dimension reduces variance, the marginal benefits of increasing $I$ and $S$ differ substantially. Increasing the number of tests $I$ consistently reduces variance. By contrast, increasing the number of partitions $S$ yields diminishing returns, with minimal improvements beyond $S = 30$ under the baseline DGP. This pattern reflects fundamental differences between additional tests and additional partitions: while increasing the number of tests results in new, independent information about performance, increasing the number of partitions eventually just provides different realizations of the same underlying information. We present results for alternative specifications of the DGP in appendix \ref{sec:supplemental_results}.

\section{Takeaways for Practitioners} 

We conclude by distilling our theoretical and simulation findings into five recommendations for sample-splitting practitioners: 

\begin{enumerate}
   \item \emph{Understand the fundamental tradeoff:} Sample splitting enables performance evaluation when true impacts are unobserved, but this comes at a cost. Because methodologies are implemented on only a split of the data, sample-split estimates are generally biased for what we ultimately care about --- how the methodologies perform on the full sample. While this bias can be reduced by increasing the split fraction $\alpha$, it cannot be eliminated entirely through the choice of split fraction alone without sacrificing our ability to conduct valid inference.
    \item \emph{Split wisely, not necessarily evenly:} The standard split fraction of $\alpha = 0.5$ may be suboptimal. Choosing a split fraction $\alpha > 0.5$ can reduce bias in sample-split performance estimates without an offsetting increase in their variance. 
    \item \emph{Skip unit-level partitioning for plug-in methods:} If you are evaluating a plug-in methodology that applies functions to the estimates $(\hat{\delta}_{ias}, \tau_{ias}^2)$ without changing the way in which they are constructed, you may not need to repeatedly partition the unit-level data underlying each A/B test. Instead, you may be able to draw sample-split estimates from their joint distribution given the full-sample estimates, as suggested in proposition 1. This can significantly lower the computational costs of sample splitting, enabling you to increase the scale of your analyses by orders of magnitude and obtain significantly more precise estimates of performance. 
    \item \emph{Be careful when the number of tests $I$ is small}: Our theoretical results are valid in the limit as the number of tests $I$ grows large. When $I$ is small, asymptotic approximations may perform poorly, resulting in unreliable estimates of performance and confidence intervals with poor coverage. Even when $I$ is sufficient for reliable asymptotic approximations, you may still lack power to detect differences in performance between methodologies. In this case, alternative approaches to performance evaluation (such as simulation) may be more appropriate. 
   \item \emph{Don't waste time and compute --- $S= 30$ is often sufficient:} Using multiple partitions reduces variance, but most of the gains accrue across small $S$, leaving relatively little benefit to $S > 30$. Unlike increasing the number of A/B tests ($I$), which brings new independent information to estimation, increasing the number of partitions ($S$) eventually just provides different realizations of the same underlying information. 

\end{enumerate}

A natural question for future work is whether the bias in takeaway 1 is unavoidable, or whether it only appears that way because the split fraction $\alpha$ is the only lever we considered for addressing it. One promising alternative lever is to reweight sample-split data so that the distribution of its sampling variances matches that of the full sample, directly correcting the source of the bias rather than trading it off against variance. We leave a full treatment of this reweighting approach, and a characterization of when it can and cannot be applied, to future work.

\bibliographystyle{aer}
\bibliography{bibliography}

@inproceedings{deng_etal_2013,
    author = {Deng, Alex and Xu, Ya and Kohavi, Ron and Walker, Toby},
    title = {Improving the sensitivity of online controlled experiments by utilizing pre-experiment data},
    year = {2013},
    isbn = {9781450318693},
    publisher = {Association for Computing Machinery},
    address = {New York, NY, USA},
    url = {https://doi.org/10.1145/2433396.2433413},
    doi = {10.1145/2433396.2433413},
    booktitle = {Proceedings of the Sixth ACM International Conference on Web Search and Data Mining},
    pages = {123–132},
    numpages = {10},
    location = {Rome, Italy},
    series = {WSDM '13}
}

@book{casella_berger_2002,
    author = {Casella, George and Berger, Roger L.},
    publisher = {Wadsworth Group},
    title = {Statistical Inference},
    year = {2002}
}

@article{azevedo_etal_2020,
    author = {Azevedo, Eduardo M. and Deng, Alex and Montiel Olea, Jos\'{e} Luis and Rao, Justin and Weyl, E. Glen},
    title = {{A/B} Testing with Fat Tails},
    journal = {Journal of Political Economy},
    volume = {128},
    number = {12},
    pages = {4614-000},
    year = {2020},
    doi = {10.1086/710607},
    URL = {https://doi.org/10.1086/710607}
}

@article{athey_imbens_2016,
    author = {Susan Athey  and Guido Imbens },
    title = {Recursive partitioning for heterogeneous causal effects},
    journal = {Proceedings of the National Academy of Sciences},
    volume = {113},
    number = {27},
    pages = {7353-7360},
    year = {2016},
    doi = {10.1073/pnas.1510489113},
    URL = {https://www.pnas.org/doi/abs/10.1073/pnas.1510489113},
    eprint = {https://www.pnas.org/doi/pdf/10.1073/pnas.1510489113}
}

@misc{tripuraneni_etal_2023,
      title={Meta-Analysis of Randomized Experiments with Applications to Heavy-Tailed Response Data}, 
      author={Nilesh Tripuraneni and Dhruv Madeka and Dean Foster and Dominique Perrault-Joncas and Michael I. Jordan},
      year={2023},
      eprint={2112.07602},
      archivePrefix={arXiv},
      primaryClass={stat.ME},
      url={https://arxiv.org/abs/2112.07602}
}

@misc{parikh_etal_2026,
      title={Towards Optimal Estimators for Randomized Control Trials},
      author={Harsh Parikh and Gabriel Levin-Konigsberg and Nilesh Tripuraneni and Dhruv Madeka and Michael I. Jordan and Dean Foster and Dominique Perrault-Joncas and Alexander Volfovsky},
      year={2026},
      eprint={2607.23254},
      archivePrefix={arXiv},
      primaryClass={stat.AP},
      url={https://arxiv.org/abs/2607.23254}
}

@article{azevedo_etal_2019,
    Author = {Azevedo, Eduardo M. and Deng, Alex and Montiel Olea, José L. and Weyl, E. Glen},
    Title = {Empirical Bayes Estimation of Treatment Effects with Many {A/B} Tests: An Overview},
    Journal = {AEA Papers and Proceedings},
    Volume = {109},
    Year = {2019},
    Month = {May},
    Pages = {43–47},
    DOI = {10.1257/pandp.20191003},
    URL = {https://www.aeaweb.org/articles?id=10.1257/pandp.20191003}
}

@misc{goldberg_johndrow_2017,
      title={A Decision Theoretic Approach to {A/B} Testing}, 
      author={David Goldberg and James E. Johndrow},
      year={2017},
      eprint={1710.03410},
      archivePrefix={arXiv},
      primaryClass={math.ST},
      url={https://arxiv.org/abs/1710.03410}
}

@inproceedings{deng_etal_2023,
    author = {Deng, Alex and Yuan, Lo-Hua and Kanai, Naoya and Salama-Manteau, Alexandre},
    title = {Zero to Hero: Exploiting Null Effects to Achieve Variance Reduction in Experiments with One-sided Triggering},
    year = {2023},
    isbn = {9781450394079},
    publisher = {Association for Computing Machinery},
    address = {New York, NY, USA},
    url = {https://doi.org/10.1145/3539597.3570413},
    doi = {10.1145/3539597.3570413},
    booktitle = {Proceedings of the Sixteenth ACM International Conference on Web Search and Data Mining},
    pages = {823–831},
    numpages = {9},
    location = {Singapore, Singapore},
    series = {WSDM '23}
}

@inproceedings{deng_etal_2024,
    author = {Deng, Alex and Hagar, Luke and Stevens, Nathaniel T. and Xifara, Tatiana and Gandhi, Amit},
    title = {Metric Decomposition in {A/B} Tests},
    year = {2024},
    isbn = {9798400704901},
    publisher = {Association for Computing Machinery},
    address = {New York, NY, USA},
    url = {https://doi.org/10.1145/3637528.3671556},
    doi = {10.1145/3637528.3671556},
    booktitle = {Proceedings of the 30th ACM SIGKDD Conference on Knowledge Discovery and Data Mining},
    pages = {4885–4895},
    numpages = {11},
    location = {Barcelona, Spain},
    series = {KDD '24}
}

@article{lewis_2025,
    author = {Lewis, Randall},
    journal = {Internal Amazon Working Paper},
    title = {Doppelgangers v. Twins for Out-of-Sample Evaluation},
    year = {2025}
}

@article{stone_1971,
    author = {Stone, M},
    journal = {Journal of the Royal Statistical Society: Series B (Methodological)},
    title = {Cross-Validatory Choice and Assessment of Statistical Predictions},
    volume = {36},
    number = {2},
    pages = {111 - 147},
    year = {1971}
}

@article{chernozhukov_2025,
    author = {Chernozhukov, Victor and Demirer, Mert and Duflo, Esther and Fernández-Val, Iván},
    title = {Fisher–Schultz Lecture: Generic Machine Learning Inference on Heterogeneous Treatment Effects in Randomized Experiments, With an Application to Immunization in India},
    journal = {Econometrica},
    volume = {93},
    number = {4},
    pages = {1121-1164},
    doi = {https://doi.org/10.3982/ECTA19303},
    url = {https://onlinelibrary.wiley.com/doi/abs/10.3982/ECTA19303},
    eprint = {https://onlinelibrary.wiley.com/doi/pdf/10.3982/ECTA19303},
    year = {2025}
}

@misc{delriochanona_etal_2023,
    title={Simulating Human Behavior with {AI} Agents},
    author={Del Rio-Chanona, R. Maria and Raman, Akhil and Srikant, Midhun and Askell, Amanda and Bai, Yuntao and Bengio, Yoshua and Brundage, Miles and Clark, Jack and Gerrish, Sean and Hadfield, Gillian and Kasirzadeh, Atoosa and Krakovna, Victoria and Krueger, David and Leike, Jan and Miller, James and Mishra, Saurabh and Sastry, Gauri and Sutton, Richard and Toner, Helen and Askell, Amanda},
    institution={Stanford Institute for Human-Centered Artificial Intelligence},
    year={2023},
    type={Policy Brief},
    url={https://hai.stanford.edu/assets/files/hai-policy-brief-simulating-human-behavior-with-ai-agents.pdf}
}

@misc{delriochanona_etal_2025,
      title={Can Generative {AI} Agents Behave Like Humans? {Evidence} from Laboratory Market Experiments},
      author={R. Maria del Rio-Chanona and Marco Pangallo and Cars Hommes},
      year={2025},
      eprint={2505.07457},
      archivePrefix={arXiv},
      primaryClass={econ.GN},
      url={https://arxiv.org/abs/2505.07457}
}

\clearpage
\appendix

\section{Appendix: Proof of Claims}
\label{sec:proofs}

\noindent \textbf{Proposition 1 (Evaluating plug-in methodologies without unit-level partitioning)}: Let splits $a$ and $b$ be formed by randomly partitioning units in an A/B test with proportions $\alpha$ and $(1-\alpha)$, respectively. Under equation (\ref{eq:likelihood}), the joint distribution of $(\hat{\Delta}_{ias}, \hat{\Delta}_{ibs})$ conditional on the full-sample estimates $(\hat{\delta}_i, \tau_i^2)$ and true impact $\delta_i$ is given by:

\begin{equation}
\begin{pmatrix}
\hat{\Delta}_{ias} \\
\hat{\Delta}_{ibs}
\end{pmatrix}
\Big |
\hat{\Delta}_i = \hat{\delta}_i, \Delta_i = \delta_i, \tau_i^2
\sim N
\begin{pmatrix}
\begin{pmatrix}
\hat{\delta}_i \\
\hat{\delta}_i
\end{pmatrix}
,
\begin{pmatrix}
\frac{\tau_i^2}{\alpha} - \tau_i^2 & -\tau_i^2 \\
-\tau_i^2 & \frac{\tau_i^2}{1-\alpha} - \tau_i^2
\end{pmatrix}
\end{pmatrix}
\end{equation}

\noindent For a given test, the process of repeatedly partitioning units and re-estimating $(\hat{\delta}_{ias}, \hat{\delta}_{ibs})$ can therefore be approximated by resampling from the conditional distribution in equation (\ref{eq:prop1}) given the full-sample estimates $(\hat{\delta}_i, \tau_i^2)$. Given draws from this distribution, plug-in methodologies (being deterministic functions of $(\hat{\delta}_{ias}, \tau_{ias}^2)$) can then be applied, yielding estimates $\{\kappa(\hat{x}_{ias}), \hat{\delta}_{ias}, \tau_{ias}^2, \hat{\delta}_{ibs}, \tau_{ibs}^2\}_{s=1}^S$.

\vspace{0.1pc}

\noindent \emph{Proof:} By the central limit theorem and the random partitioning of units into disjoint splits, we have that:

\begin{equation}
\begin{pmatrix}
\hat{\Delta}_{ias} \\
\hat{\Delta}_{ibs}
\end{pmatrix}
\Big |
\Delta_i = \delta_i, \tau_i^2
\sim N
\begin{pmatrix}
\begin{pmatrix}
\delta_i \\
\delta_i
\end{pmatrix}
,
\begin{pmatrix}
\frac{\tau_i^2}{\alpha} & 0 \\
0 & \frac{\tau_i^2}{1-\alpha} 
\end{pmatrix}
\end{pmatrix}
\end{equation}

Note that $\hat{\Delta}_i = \alpha \hat{\Delta}_{ias} + (1- \alpha) \hat{\Delta}_{ibs}$ and that, for any multivariate normal vector $V$ with mean $\mu_V$ and variance $\Sigma_V$, any affine transformation $MV$ is also multivariate normal, with mean $M \mu_V$ and variance $M \Sigma_V M'$. Defining: 

\begin{equation}
M=
\begin{pmatrix}
1 & 0 \\
0 & 1 \\
\alpha & 1-\alpha
\end{pmatrix}
\end{equation}

\noindent we therefore have that:

\begin{equation}
\begin{pmatrix}
\hat{\Delta}_{ias} \\
\hat{\Delta}_{ibs} \\
\hat{\Delta}_i
\end{pmatrix}
\Big |
\Delta_i = \delta_i, \tau_i^2
\sim N
\begin{pmatrix}
\begin{pmatrix}
\delta_i \\
\delta_i \\
\delta_i
\end{pmatrix}
,
\begin{pmatrix}
\frac{\tau_i^2}{\alpha} & 0 & \tau_i^2 \\
0 & \frac{\tau_i^2}{1-\alpha} & \tau_i^2 \\
\tau_i^2 & \tau_i^2 & \tau_i^2 
\end{pmatrix}
\end{pmatrix}
\end{equation}

Finally, note that if: 

\begin{equation}
\begin{pmatrix}
V \\
W
\end{pmatrix} \Big | \Delta_i = \delta_i, \tau_i^2 
\sim N
\begin{pmatrix}
\begin{pmatrix}
\mu_V \\
\mu_W
\end{pmatrix}
,
\begin{pmatrix}
\Sigma_{VV} & \Sigma_{VW} \\
\Sigma_{WV} & \Sigma_{WW}
\end{pmatrix}
\end{pmatrix}
\end{equation}

\noindent then $V | W = w, \Delta_i = \delta_i, \tau_i^2 \sim N\left(\mu_V + \Sigma_{VW} \Sigma_{WW}^{-1}(w - \mu_W), \Sigma_{VV} - \Sigma_{VW} \Sigma_{WW}^{-1} \Sigma_{WV}\right)$. Applying this result with $V = (\hat{\Delta}_{ias}, \hat{\Delta}_{ibs})'$ and $W = \hat{\Delta}_i$ yields:

\vspace{-0.25pc}
{\scriptsize
\begin{align}
    \mu_V + \Sigma_{VW} \Sigma_{WW}^{-1}(w - \mu_W) &= 
    \begin{pmatrix}
\delta_i \\
\delta_i
\end{pmatrix} +
\begin{pmatrix}
\tau_i^2 \\
\tau_i^2
\end{pmatrix}  \left(\frac{1}{\tau_i^2}\right) (\hat{\delta}_i - \delta_i) =
\begin{pmatrix}
\hat{\delta}_i \\
\hat{\delta}_i
\end{pmatrix} \\
    \Sigma_{VV} - \Sigma_{VW} \Sigma_{WW}^{-1} \Sigma_{WV} &= 
    \begin{pmatrix}
\frac{\tau_i^2}{\alpha} & 0 \\
0 & \frac{\tau_i^2}{1-\alpha} 
\end{pmatrix} - 
\begin{pmatrix}
\tau_i^2 \\
\tau_i^2
\end{pmatrix} \left(\frac{1}{\tau_i^2}\right)
\begin{pmatrix}
\tau_i^2 & \tau_i^2
\end{pmatrix} = 
\begin{pmatrix}
\frac{\tau_i^2}{\alpha} - \tau_i^2 & -\tau_i^2 \\
-\tau_i^2 & \frac{\tau_i^2}{1-\alpha} - \tau_i^2
\end{pmatrix}
\end{align}}

\noindent which therefore establishes the claim $\hfill \square$

\vspace{0.5pc}

\noindent \textbf{Lemma 1 (Test-level unbiasedness):} Under equations (\ref{eq:sample_split_independent})  and (\ref{eq:sample_split_unbiased}) the sample-split estimators $\hat{Y}_{is}(\kappa, \alpha)$ listed in table \ref{tab:estimands_estimators} are unbiased for their corresponding sample-split estimands $Y_{i}(\kappa, \alpha)$.

\noindent \emph{Proof:} Assume that equations (\ref{eq:sample_split_independent}) and (\ref{eq:sample_split_unbiased}) hold. For bias, the claim then follows immediately from the unbiasedness of $\hat{\Delta}_{ibs}$. For squared error, we follow \citet{tripuraneni_etal_2023}, noting that:

{\footnotesize
\vspace{-0.5pc}
\begin{align}
    E_F\left(\left(\kappa\left(\hat{X}_{ias}\right) - \hat{\Delta}_{ibs}\right)^2\right) - \tau_{ibs}^2 &= E_F\left(\left(\kappa\left(\hat{X}_{ias}\right) - \delta_i + \delta_i - \hat{\Delta}_{ibs}\right)^2\right) - \tau_{ibs}^2 \\
     &= E_F\left(\left(\kappa\left(\hat{X}_{ias}\right) - \delta_i\right)^2  +  \left( \delta_i - \hat{\Delta}_{ibs}\right)^2 \right) -  \tau_{ibs}^2 \label{eq:lemma1_1} \\
     &= E_F\left(\left(\kappa\left(\hat{X}_{ias}\right) - \delta_i\right)^2 \right) \label{eq:lemma1_2} 
\end{align}}

\noindent where equation (\ref{eq:lemma1_1}) follows from the independence of $\kappa(\hat{X}_{ias})$ and $\hat{\Delta}_{ibs}$ conditional on $(\delta_i, z_i)$ and the unbiasedness of $\hat{\Delta}_{ibs}$. The claim then follows in equation (\ref{eq:lemma1_2}) from the definition of the sampling variance of $\hat{\Delta}_{ibs}$. For launch-only decision value, we follow \citet{tripuraneni_etal_2023}, noting that:

{\footnotesize
\vspace{-1pc}
\begin{align}
    E_F\left(\kappa\left(\hat{X}_{ias}\right) \cdot \hat{\Delta}_{ibs}\right) &= E_F\left( \kappa\left(\hat{X}_{ias}\right)\right) E_F\left(\hat{\Delta}_{ibs} \right)  \label{eq:lemma1_3}  \\
     &=  E_F\left( \kappa\left(\hat{X}_{ias} \right)  \cdot \delta_i \right)  \label{eq:lemma1_4} 
\end{align}}

\noindent where equations (\ref{eq:lemma1_3}) and (\ref{eq:lemma1_4})  follow from the independence of $\kappa(\hat{X}_{ias})$ and $\hat{\Delta}_{ibs}$ conditional on $(\delta_i, z_i)$ and the unbiasedness of $\hat{\Delta}_{ibs}$. The same argument applies to decision value $\hfill \square$

\vspace{0.5pc}

\noindent \textbf{Proposition 2 (Unbiasedness)}: Under equations (\ref{eq:split_estimand_eq2}) and (\ref{eq:sample_split_estimator_eq1}), the estimator of average performance $\hat{\theta}(\kappa, \alpha, S)$ is unbiased for $\theta(\kappa, \alpha)$.

\vspace{0.5pc}

\noindent \emph{Proof:} Under equations (\ref{eq:split_estimand_eq2}) and (\ref{eq:sample_split_estimator_eq1}), we have that:

\vspace{-1pc}
\begin{align}
    E\left(\hat{\theta}(\kappa, \alpha, S)\right) &= E_G\left(E_F\left(\hat{\theta}(\kappa, \alpha, S)\right)\right) \label{eq:prop2_1} \\
    &= E_G\left(E_F\left(\frac{1}{I} \sum_{i=1}^I \hat{Y}_{i}(\kappa, \alpha, S) \right)\right)  \label{eq:prop2_2}  \\
    &= E_G\left(\frac{1}{I} \sum_{i=1}^I  E_F\left(\hat{Y}_{i}(\kappa, \alpha, S) \right)\right)  \label{eq:prop2_3}  \\
    &= E_G\left(Y_i\left(\kappa, \alpha\right)\right)  \label{eq:prop2_4}  \\
    &= \theta(\kappa, \alpha)  \label{eq:prop2_5} 
\end{align}

\noindent where equation (\ref{eq:prop2_1}) follows from the law of iterated expectations, equation (\ref{eq:prop2_2}) follows from the definition of $\hat{\theta}(\kappa, \alpha, S)$, equation (\ref{eq:prop2_3}) follows from the linearity of expectations, equation (\ref{eq:prop2_4}) follows from equation (\ref{eq:sample_split_estimator_eq1}), and equation (\ref{eq:prop2_5}) follows from equation (\ref{eq:split_estimand_eq2}) $\hfill \square$

\vspace{0.5pc}

\noindent \textbf{Lemma 2 (Consistency):} Under equations (\ref{eq:split_estimand_eq2}) and (\ref{eq:sample_split_estimator_eq1}) and the regularity condition that $E(\hat{Y}_i(\kappa, \alpha, S)^2) < \infty$,  the estimator $\hat{\theta}(\kappa, \alpha, S)$ is consistent for $\theta(\kappa, \alpha)$.

\noindent \emph{Proof:} Under equations (\ref{eq:split_estimand_eq2}) and (\ref{eq:sample_split_estimator_eq1})  and the regularity condition that $E(\hat{Y}_i(\kappa, \alpha, S)^2) < \infty$:

\vspace{-1pc}
\begin{align}
    \hat{\theta}(\kappa, \alpha, S) &\overset{p}{\longrightarrow} E\left(\hat{Y}_i\left(\kappa, \alpha, S\right)\right) \label{eq:lemma2_1} \\
    &= E_G\left(E_F\left(\hat{Y}_i\left(\kappa, \alpha, S\right)\right)\right) \label{eq:lemma2_2} \\
    &= E_G\left(Y_i \left(\kappa, \alpha \right)\right)\label{eq:lemma2_3} \\
    &= \theta(\kappa, \alpha) \label{eq:lemma2_4}
\end{align}

\noindent where equation (\ref{eq:lemma2_1}) follows from the law of large numbers, equation (\ref{eq:lemma2_2}) follows from the law of iterated expectations, equation (\ref{eq:lemma2_3}) follows from equation (\ref{eq:sample_split_estimator_eq1}), and equation (\ref{eq:lemma2_4}) follows from equation (\ref{eq:split_estimand_eq2}) $\hfill \square$

\vspace{0.5pc}

\noindent \textbf{Proposition 3 (Asymptotic normality)}: Under equations (\ref{eq:split_estimand_eq2}) and (\ref{eq:sample_split_estimator_eq1}) and the regularity condition that $E(\hat{Y}_i(\kappa, \alpha, S)^2) < \infty$, the estimator $\hat{\theta}(\kappa, \alpha, S)$ is asymptotically normally distributed:

\begin{equation}
    \sqrt{I}\left(\hat{\theta}(\kappa, \alpha, S) - \theta(\kappa, \alpha)\right) \overset{d}{\longrightarrow} N\left(0, \zeta^2(\kappa, \alpha, S) \right)
\end{equation}

\noindent where $\zeta^2(\kappa, \alpha, S) =  \lambda^2(\kappa, \alpha) + E_G\left(\Omega_i^2\left(\kappa, \alpha, S\right)\right)$.

\noindent \emph{Proof:} First, note that:

\vspace{-1pc}
{\footnotesize 
\begin{align}
    \text{Var}\left(\hat{Y}_i\left(\kappa, \alpha, S\right)\right) &= \text{Var}_G\left(E_F\left(\hat{Y}_{i}\left(\kappa, \alpha, S\right)\right)\right) + E_G\left(\text{Var}_F\left(\hat{Y}_{i}\left(\kappa, \alpha, S\right)\right)\right) \label{eq:prop3_1} \\
    &= \text{Var}_G\left(Y_i\left(\kappa, \alpha\right)\right) + E_G\left(\Omega_i^2\left(\kappa, \alpha, S\right)\right)  \label{eq:prop3_2} \\
    &= \lambda^2(\kappa, \alpha) + E_G\left(\Omega_i^2\left(\kappa, \alpha, S\right)\right)  \label{eq:prop3_3}  
\end{align}
}%

\noindent where equation (\ref{eq:prop3_1}) follows from the law of total variance, equation (\ref{eq:prop3_2}) follows from equation (\ref{eq:sample_split_estimator_eq1}), and equation (\ref{eq:prop3_3}) follows from equation (\ref{eq:split_estimand_eq2}).

Given that $\hat{Y}_i(\kappa, \alpha, S)$ are independently and identically distributed across tests and the regularity condition that $E(\hat{Y}_i(\kappa, \alpha, S)^2) < \infty$, the result then follows from the central limit theorem $\hfill \square$

\vspace{0.5pc}

\noindent \textbf{Proposition 4 (Valid asymptotic inference):} Under equations (\ref{eq:split_estimand_eq2}) and (\ref{eq:sample_split_estimator_eq1}) and the regularity condition that $E(\hat{Y}_i(\kappa, \alpha, S)^2) < \infty$, the estimator:

\begin{equation}
 \hat{\zeta}^2(\kappa, \alpha, S) = \frac{1}{I} \sum_{i=1}^I (\hat{Y}_i(\kappa, \alpha, S) - \hat{\theta}(\kappa, \alpha, S))^2
\end{equation}
 
\noindent is consistent for the asymptotic variance $\zeta^2(\kappa, \alpha, S) =  \lambda^2(\kappa, \alpha) + E_G\left(\Omega_i^2\left(\kappa, \alpha, S\right)\right)$. Therefore, the $(1-\upsilon)$ confidence intervals:

\begin{equation}
    \hat{\theta}(\kappa, \alpha, S) \pm \frac{\hat{\zeta}(\kappa, \alpha, S)}{\sqrt{I}} \cdot \Phi^{-1}\left(1-\frac{\upsilon}{2}\right) \\
\end{equation}

\noindent will cover $\theta(\kappa, \alpha)$ with probability $(1-\upsilon)$ as $I \longrightarrow \infty$.

\noindent \emph{Proof:} We can decompose the estimator $\hat{\zeta}^2(\kappa, \alpha, S)$ as follows:

\vspace{-1pc}
{\footnotesize
\begin{align}
    \hat{\zeta}^2(\kappa, \alpha, S) &= \frac{1}{I} \sum_{i=1}^I \left( \left(\hat{Y}_i - \theta \right)^2 - 2 \left(\hat{Y}_i - \theta \right)\left(\hat{\theta}- \theta \right) + (\hat{\theta}- \theta)^2 \right) \\
    &= \frac{1}{I} \sum_{i=1}^I \left(\hat{Y}_i - \theta\right)^2 - \frac{2 \left(\hat{\theta} - \theta\right)}{I}\sum_{i=1}^I \left(\hat{Y}_i - \theta\right) + (\hat{\theta} - \theta)^2 \\ 
    &= \frac{1}{I} \sum_{i=1}^I \left(\hat{Y}_i - \theta\right)^2 - \left(\hat{\theta} - \theta\right)^2
\end{align}
}%

\noindent where we suppress the $(\kappa, \alpha, S)$ and $(\kappa, \alpha)$ notation in the above for simplicity. Under equations (\ref{eq:split_estimand_eq2}) and (\ref{eq:sample_split_estimator_eq1}) and the regularity condition that $E(\hat{Y}_i(\kappa, \alpha, S)^2) < \infty$, the first term converges in probability to $\zeta^2(\kappa, \alpha, S)$ by the law of large numbers. The claim that $\hat{\zeta}^2(\kappa, \alpha, S) \overset{p}{\longrightarrow} \zeta^2(\kappa, \alpha, S)$ then follows from lemma 2 and the continuous mapping theorem, which together imply that the second term $(\hat{\theta}(\kappa, \alpha, S) - \theta(\kappa, \alpha))^2$ converges in probability to zero. Given this result, Slutsky's theorem then implies that:

\begin{equation}
    \frac{\sqrt{I}\left(\hat{\theta}(\kappa, \alpha, S) - \theta(\kappa, \alpha)\right)}{\hat{\zeta}(\kappa, \alpha, S)} \overset{d}{\longrightarrow} N\left(0, 1\right)
\end{equation}

\noindent from which the claim regarding confidence intervals follows immediately $\hfill \square$

\clearpage

\section{Appendix: Supplemental Results}
\label{sec:supplemental_results}

\begin{figure}[h!]
\centering 
\caption{Bias-variance tradeoff underlying choice of split fraction $\alpha$ (robustness)}
\begin{tabular}{cc}
    \multicolumn{2}{c}{(a) $(\sigma^2, \tau^2) = (1, 4)$} \\ 
    {\scriptsize (i)  Bayes estimator relative to unbiased } & {\scriptsize (ii) Bayes decision rule relative to frequentist}  \\ [-0.3pc]
    {\scriptsize estimator under squared error} & {\scriptsize decision rule under launch-only decision value} \\ 
     \includegraphics[scale = 0.47]{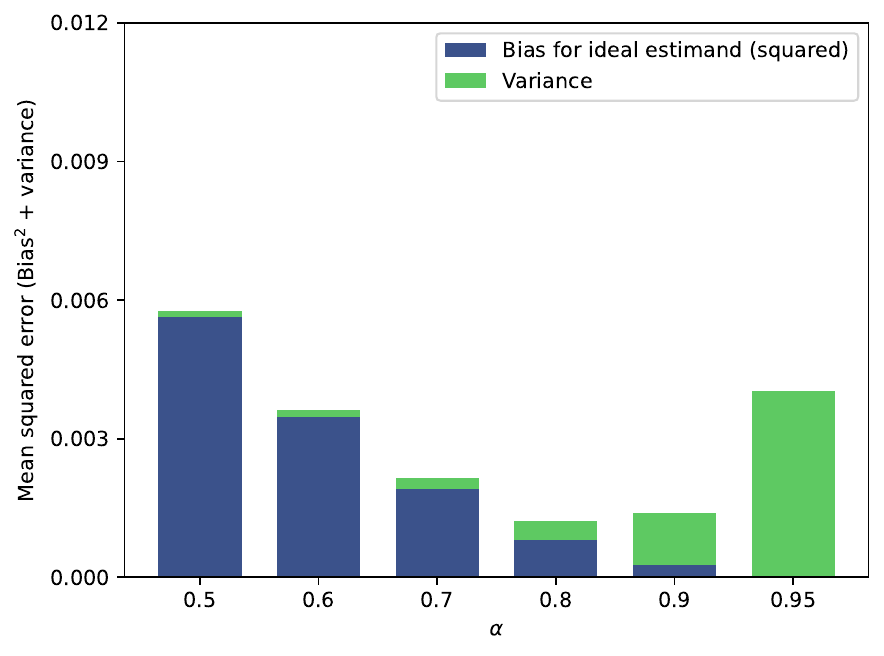} & \includegraphics[scale = 0.47]{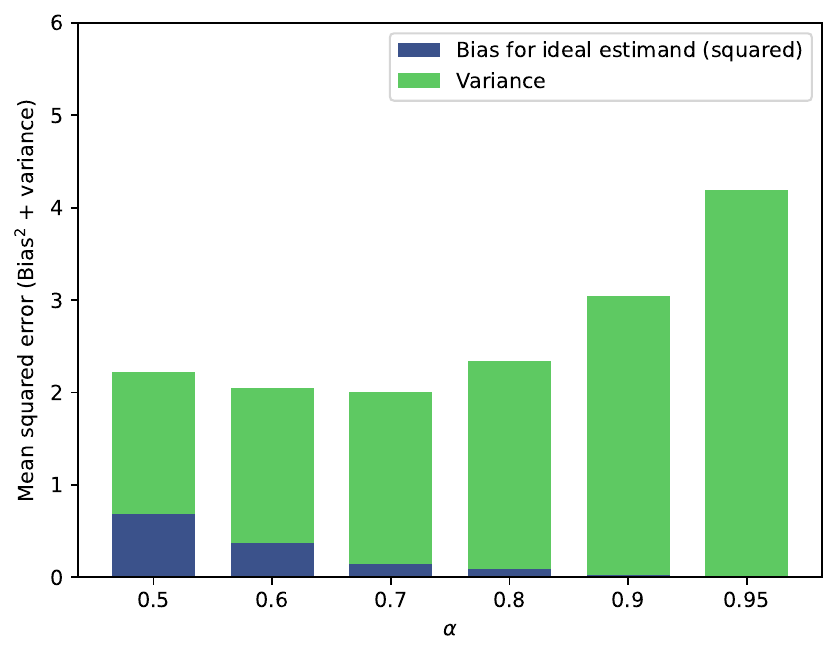} \\ [0.25pc]
     \multicolumn{2}{c}{(b)  $(\sigma^2, \tau^2) = (1, 8)$} \\ 
    {\scriptsize (i)  Bayes estimator relative to unbiased } & {\scriptsize (ii) Bayes decision rule relative to frequentist}  \\ [-0.3pc]
    {\scriptsize estimator under squared error} & {\scriptsize decision rule under launch-only decision value} \\ 
     \includegraphics[scale = 0.47]{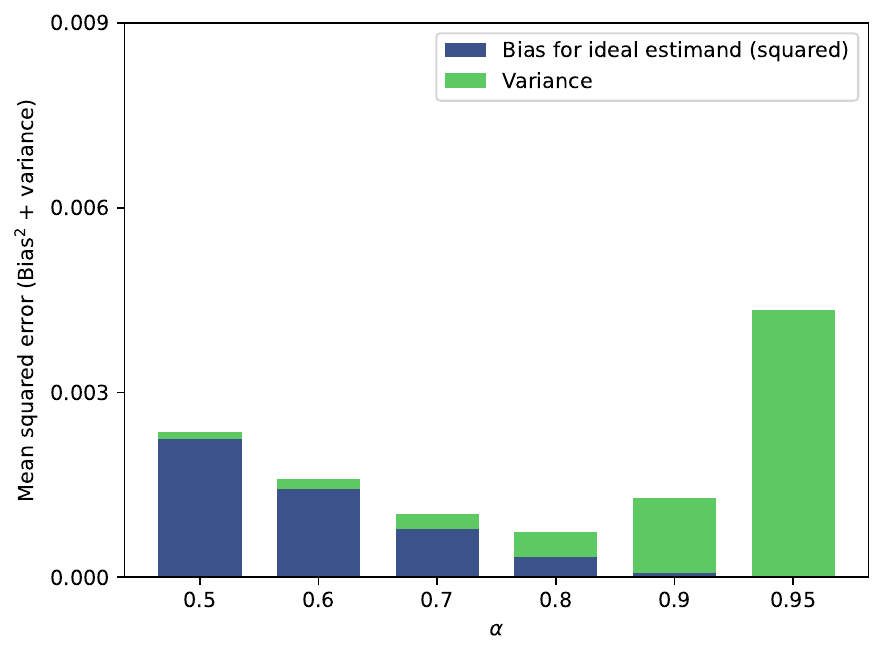} & \includegraphics[scale = 0.47]{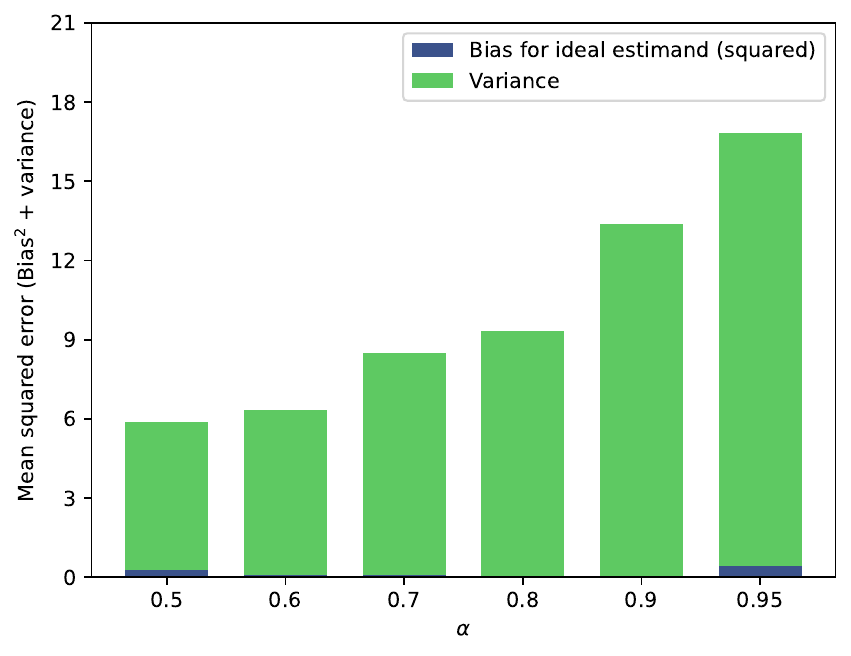} \\ [0.25pc]
\end{tabular}
\end{figure}

\begin{figure}[h!]
\centering 
\caption{Variance as a function of the number of tests $I$ and number of partitions $S$ (robustness)}
\begin{tabular}{cc}
    \multicolumn{2}{c}{(a) $(\sigma^2, \tau^2) = (1, 4)$} \\ 
    {\scriptsize (i)  Bayes estimator relative to unbiased } & {\scriptsize (ii) Bayes decision rule relative to frequentist}  \\ [-0.3pc]
    {\scriptsize estimator under squared error} & {\scriptsize decision rule under launch-only decision value} \\ 
     \includegraphics[scale = 0.47]{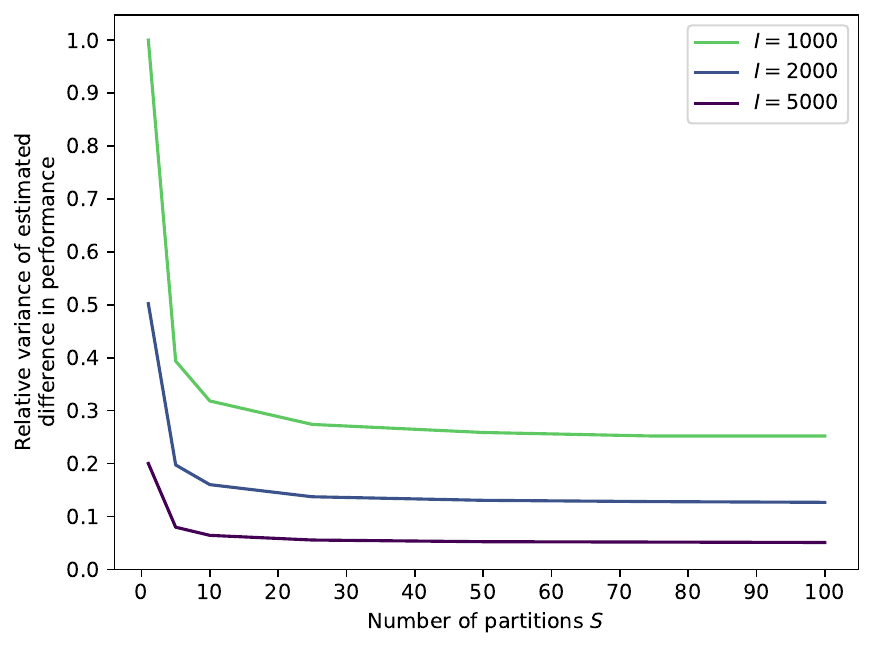} & \includegraphics[scale = 0.47]{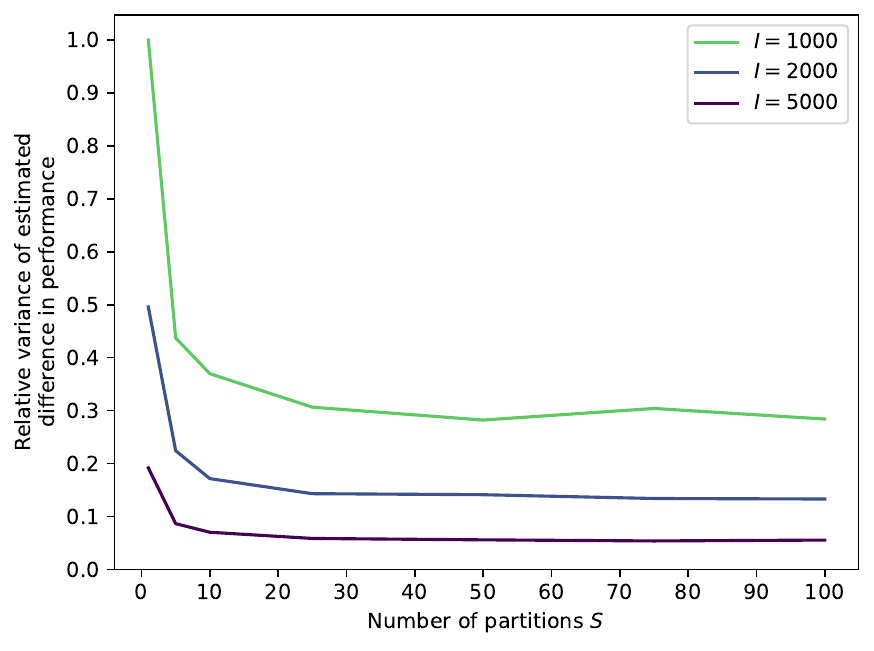} \\ [0.25pc]
     \multicolumn{2}{c}{(b) $(\sigma^2, \tau^2) = (1, 8)$} \\ 
    {\scriptsize (i)  Bayes estimator relative to unbiased } & {\scriptsize (ii) Bayes decision rule relative to frequentist}  \\ [-0.3pc]
    {\scriptsize estimator under squared error} & {\scriptsize decision rule under launch-only decision value} \\ 
     \includegraphics[scale = 0.47]{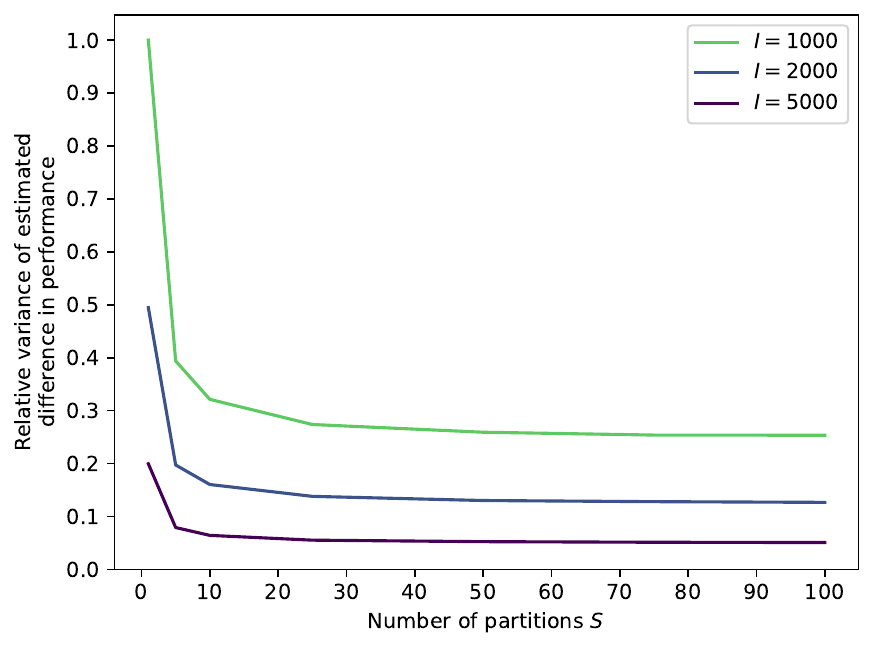} & \includegraphics[scale = 0.47]{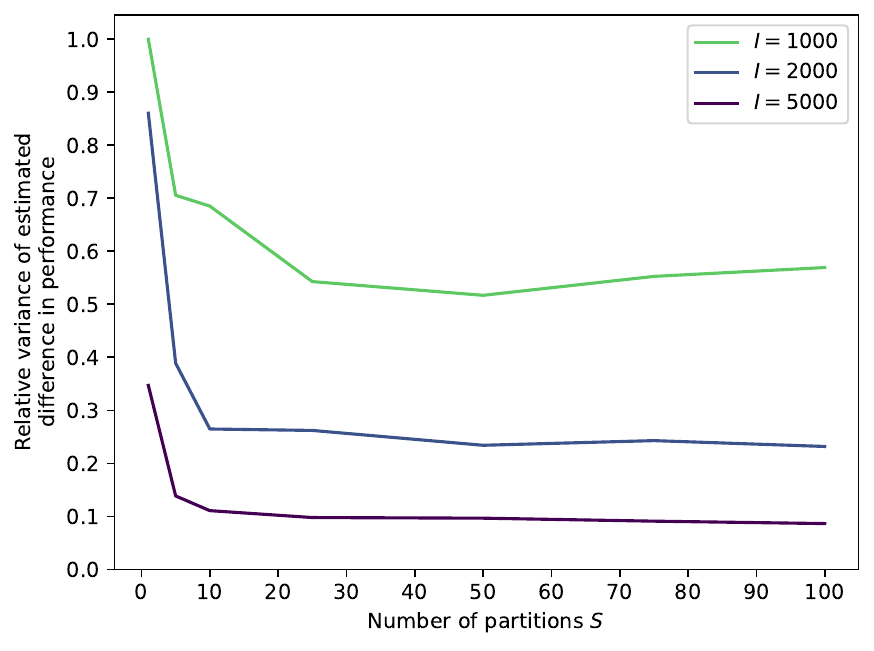} \\ [0.25pc]
     \multicolumn{2}{c}{(c) $\alpha = 0.8$} \\ 
     {\scriptsize (i)  Bayes estimator relative to unbiased } & {\scriptsize (ii) Bayes decision rule relative to frequentist}  \\ [-0.3pc]
    {\scriptsize estimator under squared error} & {\scriptsize decision rule under launch-only decision value} \\ 
     \includegraphics[scale = 0.47]{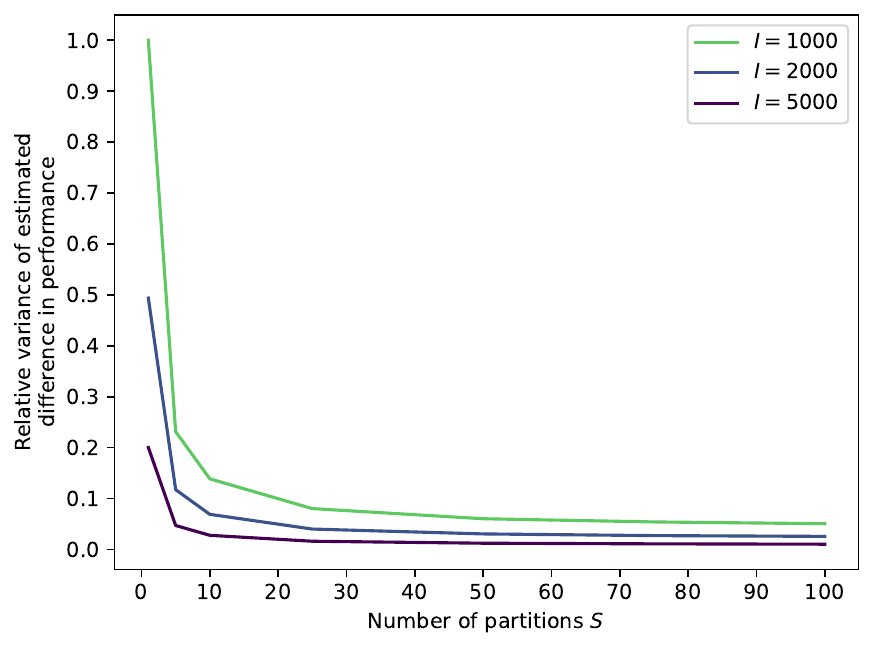} & \includegraphics[scale = 0.47]{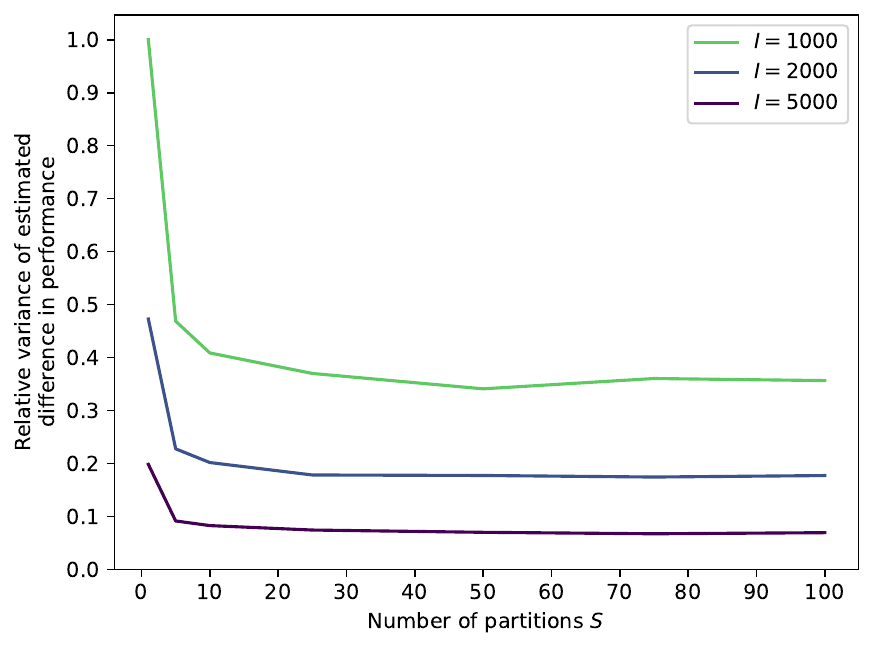} \\ [0.25pc]
\end{tabular}
\end{figure}

\end{document}